\documentclass[sigconf, preprint]{acmart}

\usepackage{graphicx}
\usepackage{subcaption}
\usepackage{todonotes}
\usepackage{amsmath}
\usepackage{booktabs}
\usepackage{siunitx}
\usepackage{threeparttable}
\usepackage{longtable}

\usepackage{geometry}
\AtBeginDocument{%
  }

\renewcommand\footnotetextcopyrightpermission[1]{}  %
\acmConference{}{}{}
\acmBooktitle{}
\begin{document}

\title{The Digital Divide in Generative AI: Evidence from Large Language Model Use in College Admissions Essays}


\author{Jinsook Lee}
\affiliation{%
  \institution{Cornell University}
  \city{Ithaca}
  \country{USA}}
\email{jl3369@cornell.edu}

\author{Conrad Borchers}
\affiliation{%
  \institution{Carnegie Mellon University}
  \city{Pittsburgh}
  \country{USA}}
\email{cborcher@cs.cmu.edu}

\author{AJ Alvero}
\affiliation{%
  \institution{Cornell University}
  \city{Ithaca}
  \country{USA}}
\email{ajalvero@cornell.edu}

\author{Thorsten Joachims}
\affiliation{%
  \institution{Cornell University}
  \city{Ithaca}
  \country{USA}}
\email{thorsten.joachims@cornell.edu}

\author{Rene F. Kizilcec}
\affiliation{%
  \institution{Cornell University}
  \city{Ithaca}
  \country{USA}}
\email{kizilcec@cornell.edu}
\renewcommand{\shortauthors}{Lee et al.}

\begin{abstract}
Large language models (LLMs) have become popular writing tools among students and may expand access to high-quality feedback for students with less access to traditional writing support. At the same time, LLMs may standardize student voice or invite overreliance. This study examines how adoption of LLM-assisted writing varies across socioeconomic groups and how it relates to outcomes in a high-stakes context: U.S. college admissions. We analyze a de-identified longitudinal dataset of applications to a selective university from 2020 to 2024 (N = 81,663). Estimating LLM use using a distribution-based detector trained on synthetic and historical essays, we tracked how student writing changed as LLM use proliferated, how adoption differed by socioeconomic status (SES), and whether potential benefits translated equitably into admissions outcomes. Using fee-waiver status as a proxy for SES, we observe post-2023 convergence in surface-level linguistic features, with the largest changes in fee-waived and rejected applicants. 
Estimated LLM use rose sharply in 2024 across all groups, with disproportionately larger increases among lower SES applicants,  consistent with an access hypothesis in which LLMs substitute for scarce writing support. However, increased estimated LLM use was more strongly associated with declines in predicted admission probability for lower SES applicants than for higher SES applicants, even after controlling for academic credentials and stylometric features. These findings raise concerns about equity and the validity of essay-based evaluation in an era of AI-assisted writing and provide the first large-scale longitudinal evidence linking LLM adoption, linguistic change, and evaluative outcomes in college admissions.
\end{abstract}



\maketitle

\section{Introduction}

The digital divide has historically been framed as a gap in access to technology. Subsequent scholarship has shown that even when access is universal, differences in use and context shape who ultimately benefits \cite{warschauer2004technology}. Large language models (LLMs) introduce a new chapter in this trajectory. These tools are widely accessible and rapidly adopted, yet concerns are growing that differences in use and evaluation may produce unequal benefits across student populations \cite{viberg2024advancing,yan2024practical}.

This tension is especially salient in college admissions essays, a high-stakes writing context where students are evaluated on their ability to communicate experiences, perspectives, and writing ability beyond quantitative metrics of college aptitude \cite{warren2013rhetoric, early2011making, early2010write, jones2013ensure, aukerman2018student}. The emergence of LLMs has the potential to reshape how these essays are produced. LLMs can support brainstorming, drafting, paraphrasing, and editing, lowering barriers to writing and potentially expanding access to high-quality feedback \cite{early2011making, polakova2024impact, khampusaen2025impact, zdravkova2025impact, korchak2025enhancing}. For students with limited access to tutors, essay coaches, or well-funded writing centers, these tools may function as substitutes for scarce writing support.

At the same time, widespread LLM use raises concerns about over-reliance, homogenization of writing, and shifting evaluative meaning. Prior work suggests that AI-generated text converges toward common linguistic patterns and rhetorical structures \cite{moon2024homogenizing}. In admissions contexts, where essays are intended to differentiate applicants, this convergence may reduce the signal value of writing and complicate how essays are interpreted by readers \cite{alvero2021essay}. Moreover, because LLM outputs reflect patterns in training data that overrepresent highly educated writers, AI-assisted essays may reproduce linguistic norms associated with privilege \cite{alvero2024large, lee2025poor}.

Emerging evidence suggests that LLM adoption and use may differ across socioeconomic groups. For instance, Bassignana et al.\cite{bassignana2025ai} found that higher SES students tend to employ more abstract and sophisticated prompting strategies, while lower SES students use more concrete prompts. Although both groups have relatively easy access to LLMs, the gains in writing, such as improved grammar and other qualities, appear more concentrated among higher SES students \cite{yu2024whose}. This raises an important question: if access to AI writing support is widespread, will differences in use and evaluation lead to unequal academic outcomes? In other words, LLMs may reduce disparities in access to support while simultaneously introducing new disparities in how that support translates into success.

Despite growing attention to AI-assisted writing, we lack large-scale empirical evidence on how writing practices have changed since the public release of ChatGPT, whether adoption varies by socioeconomic status at scale, and whether the relationship between essay characteristics and admissions outcomes has shifted in the era of AI-mediated writing. These questions are particularly urgent in college admissions, where essays play a central role in high-stakes decisions.

This study addresses these gaps using a large-scale longitudinal dataset of college admission essays submitted to a highly selective U.S. university between 2020 and 2024. We pursue three research questions: (1) How have linguistic features of admission essays changed over time across socioeconomic and admissions groups? (2) How has LLM-assisted writing adoption evolved, and does it differ by socioeconomic status? (3) Has the relationship between essay characteristics, estimated LLM use, and predicted admission probability shifted following widespread adoption of LLMs?

This paper makes three contributions. First, we provide a large-scale longitudinal analysis of college admission essays spanning the pre- and post-ChatGPT era, documenting large scale changes in linguistic features across socioeconomic and admissions groups. Second, we introduce and validate a distribution-based approach (inspired by \cite{liangmonitoring}) for estimating LLM-assisted writing at the document level, enabling analysis of adoption patterns at scale in the absence of ground-truth usage data. Third, we link estimated LLM adoption to evaluative outcomes, showing that increased use is associated with differential admissions penalties across socioeconomic groups even after controlling for academic credentials and stylometric features. Together, these contributions provide empirical evidence on how generative AI may reshape equity in high-stakes writing and offer a framework for studying AI adoption and outcomes in other educational contexts.

\section{Related Work}

\subsection{College Admission Essays and Socioeconomic Groups}

Admission essays play a central role in selective college admissions. However, prior work shows strong relationships between essay content, style, and family income, with higher-income students producing essays with greater lexical diversity and syntactic complexity \cite{alvero2021essay}. These disparities stem partly from unequal access to writing support resources: professional essay coaching and editing services are highly stratified by socioeconomic status, allowing affluent families to purchase specialized guidance that shapes application narratives \cite{huang2024translating}.

Inequality also reflects differences in cultural capital and familiarity with dominant academic discourse patterns \cite{nash1990bourdieu, jaeger2018cultural}. Students from higher SES backgrounds often possess greater familiarity with the rhetorical conventions valued in selective admissions, accumulated through years of exposure to academic writing and college-preparatory environments. In contrast, lower SES students may face disadvantages when inferring admissions expectations and crafting narratives aligned with institutional norms \cite{warren2013rhetoric, cho2025academic}.

The evaluation process itself may further compound these disparities. Experimental evidence suggests that admissions readers do not evaluate essay characteristics in isolation but rather contextualize them based on other applicant information \cite{bastedo2022contextualizing}. Ethnographic research reveals that readers attend to signals of authenticity and may view highly polished essays with suspicion, particularly when they appear inconsistent with other aspects of an applicant's profile \cite{stevens2009creating, huang2024translating}. This creates a paradox: students who receive extensive coaching may produce higher-quality essays yet face skepticism about whether those essays reflect their genuine voice, while students lacking access to support may submit essays that fail to meet evaluative standards for linguistic sophistication. The result is a system where socioeconomic advantage operates through multiple, sometimes contradictory mechanisms. Access to coaching improves essay quality but may simultaneously trigger concerns about authenticity and outside assistance. This dynamic suggests that external assistance may be interpreted differently depending on an applicant's background, which becomes especially relevant in the context of AI-assisted writing.

\subsection{Risks of LLM-Assisted Writing}

While LLM assistance can enhance writing productivity and quality, particularly among less experienced writers \cite{noy2023experimental}, and improve dimensions such as grammar, vocabulary, and organizational coherence \cite{polakova2024impact, kim2023towards}, these benefits depend critically on how the tools are deployed. Evidence suggests that students can over-rely on LLM suggestions without deeper engagement \cite{myung2025scaffolding, deng2025does, stofiana2025writing}, and that heavy use may reduce cognitive investment in writing \cite{kosmyna2025your}. Beyond individual-level concerns, researchers have documented troubling patterns of homogenization in AI-assisted writing at the collective level. Studies have found that LLM-generated text tends to converge on similar linguistic patterns, rhetorical moves, and stylistic choices that reflect the statistical regularities in models' training data rather than individual writers' authentic voices \cite{anderson2024homogenization}. While LLM assistance may boost individual creativity in some contexts, it simultaneously reduces collective novelty and diversity across a population of writers \cite{doshi2024generative}. This homogenization effect is particularly concerning given evidence that AI writing suggestions push text toward Western stylistic norms, potentially marginalizing non-Western rhetorical traditions and discourse patterns \cite{agarwal2025ai, zhang2025generative}.

For college admissions, where essays have historically served to surface individual experiences and distinctive perspectives, this convergence toward standardized linguistic patterns poses significant challenges. Research has documented how AI tools can diminish the authenticity and voice that distinguish individual writers \cite{hwang202580}. When thousands of applicants use similar LLM tools to refine their essays, the resulting texts may become increasingly difficult to differentiate, undermining the very purpose of personal statements as mechanisms for holistic evaluation. This tension between polish and authenticity creates new dilemmas for both students, who must decide how much AI assistance to use, and admissions officers, who must interpret writing in an era where surface-level quality may not reliably signal underlying ability or genuine voice.

\subsection{Socioeconomic Differences in LLM Adoption and Use}

While LLMs are widely accessible, emerging evidence suggests that students from different socioeconomic backgrounds adopt and use them differently. Skills for eliciting high-quality outputs, such as effective prompting and strategic integration of AI suggestions, appear stratified by prior educational and technological experience \cite{freeman2025student}. Studies comparing prompting patterns across socioeconomic groups has documented systematic differences: higher SES students tend to employ more abstract and sophisticated prompting strategies, while lower SES students use more concrete, straightforward prompts \cite{bassignana2025ai}.

These differences in use may translate into unequal benefits from LLM assistance. While LLM assistance can improve writing quality across groups, gains appear more concentrated among higher SES students who possess greater facility with prompt engineering and more strategic approaches to integrating AI suggestions into their writing \cite{yu2024whose}. Because LLM outputs reflect linguistic norms prevalent in their training corpora dominated by highly educated writers, AI-assisted text may inadvertently reproduce and amplify existing inequalities in cultural capital \cite{nghiem2025rich}. The result is a digital divide not merely in access to LLM tools, which are largely free and widely available, but in the ability to use those tools effectively and in ways that align with evaluative expectations in high-stakes contexts.

\vspace{-6pt}

\section{Methodology}

Our analysis proceeds in four stages. First, we construct a longitudinal dataset of admission essays and applicant characteristics spanning the pre- and post-ChatGPT era. Second, we measure changes in essay linguistic features over time (RQ1). Third, we estimate LLM-assisted writing at the document level using a distribution-based detector trained on synthetic and historical essays (RQ2). Finally, we examine how LLM adoption relates to admissions outcomes using difference-in-differences and regression models (RQ3). Details of the dataset, robustness checks, and supplementary analyses are provided in the Appendix.

\subsection{Dataset and Preprocessing}
We analyze a de-identified longitudinal dataset of applications (N = 81,663) to a selective U.S. university spanning five admission cycles from the 2019 to 2020 cycle through the 2023 to 2024 cycle. We partition the data into pre-GPT era and post-GPT era to examine changes in essay writing patterns following the widespread availability of large language models. ChatGPT was released to the public on November 30, 2022. Given that the Regular Decision application deadline for the 2023 admissions cycle was January 1, 2023 (only one month after ChatGPT's release) and considering the significant wait list demand and limited access during this initial period, we treat the 2020 to 2023 cycles as the pre-GPT era, while the 2024 cycle is treated as the post-GPT era.

Applicant SES is proxied using fee-waiver status based on documented financial need. Fee-waiver status is widely used in admissions research as a proxy for financial need and socioeconomic disadvantage. Following prior work on admissions modeling \citep{nghiem2025rich}, we create confounding variables from our dataset for applicant characteristics including sex, first-generation status, school type, scaled GPA, standardized test scores (SAT/ACT), and honors. The outcome is a binary admission decision, where admitted, conditionally admitted, and waitlisted applicants are coded as positive outcomes, reflecting advancement in the admissions process, while rejected applicants are coded as negative outcomes.

All data were de-identified prior to analysis and used under institutional data governance and research ethics approval.

\subsection{LLM Usage Estimation}
\subsubsection{Synthetic essay generation}
Because ground-truth LLM usage is unavailable in observational admissions data, we generate synthetic essays to construct reference distributions of LLM-generated text. We randomly sampled 30,000 applicants from applications submitted to the case institution over four admissions cycles (2019--2020 through 2022--2023) and analyzed their Common Application essays. We excluded essays shorter than 250 words, yielding a final dataset of 29,232 applicants and essays. We generated synthetic essays by prompting GPT-4o \cite{openai2024gpt4o} with the original Common Application essay questions. The prompt instructed the model to write as a high school student applying to the case institution's College of Engineering, following the Common App essay instructions and word limits (250--650 words). For each synthetic essay, the specific essay prompt was varied to match the distribution of prompt choices observed in the human essay sample (see Appendix~\ref{appdx:case_inst} for the full prompt template and Table~\ref{tab:llm_prompt}).

\subsubsection{Estimation of LLM-generated text}
To estimate the extent of LLM involvement in each essay, we adapted the distributional GPT quantification framework of \cite{liangmonitoring} to the essay level. The original method estimates a single mixing proportion $\alpha$ for an entire corpus by modeling it as a mixture of human-written and LLM-generated text distributions. We extend this approach to individual essays, computing an essay-level score by comparing each essay's token-level likelihood profile against reference distributions of human-written and LLM-generated text. The resulting estimate, denoted $\hat{\alpha}$, is a continuous value ranging from 0 to 1, where values near 0 indicate that the essay's distributional signature is indistinguishable from the human reference, and higher values indicate greater similarity to LLM-generated text.

For model training, we used applications from the 2019-2020 through 2022-2023 cycles, split 80:20 into training and validation sets. All essays in these pre-ChatGPT cycles serve as the human reference corpus. To construct the LLM reference corpus, we prompted an LLM to generate synthetic essays that match the demographic and prompt distributions of the human training examples. Our target dataset consisted of applications from the 2023--2024 cycle, the first full cycle following ChatGPT's public release.
Figure \ref{fig:MLE_validation} in Appendix \ref{appdx:llm_distribution_validation} validates our estimation methodology, showing strong calibration between predicted and ground-truth $\alpha$ values with minimal deviation from the identity line. We emphasize that $\hat{\alpha}$ captures relative, aggregate trends and does not constitute ground-truth identification of individual LLM use. The estimates reflect the degree to which an essay exhibits linguistic patterns characteristic of LLM-generated content, as learned by the detector from differences between human and machine writing in the reference corpora.

\subsubsection{LLM Usage Difference in SES groups in the Post-GPT Period}
Based on our observation of distribution of $\hat{\alpha}$ on the Post-GPT essays (Figure \ref{fig:alpha_distribution} Appendix \ref{appdx:llm_distribution_validation}),  we separated applicants with no detectable LLM involvement ($\hat{\alpha} = 0$). We then divided the remaining applicants into terciles based on the empirical distribution of estimated $\hat{\alpha}$, yielding three groups of approximately equal size. We then labeled the four groups: no LLM use ($\hat{\alpha} = 0$), low LLM use ($0 < \hat{\alpha} \leq 0.07$), medium LLM use ($0.07 < \hat{\alpha} \leq 0.13$), and high LLM use ($\hat{\alpha} > 0.13$). 

We then tested whether LLM usage differed significantly across SES groups in the post-GPT period. Within the post-GPT sample (N = 17,654), we examined whether LLM usage differed by socioeconomic status through both continuous and categorical analyses. First, we compared mean $\hat{\alpha}$ values across income groups using independent samples $t$-tests. Second, we examined the distribution of applicants across the four LLM usage categories (no, low, medium, high) by income level using cross-tabulation and chi-square tests. These analyses reveal whether low-income and high-income students differed in estimated LLM usage.

\subsection{Measuring Linguistic Changes in Admission Essays}
To characterize population-level changes in writing over time and explore potential mechanisms underlying outcome differences, we compute lexical diversity measures including type--token ratio (TTR) \cite{templin1957certain}, Maas TTR \cite{maas1972uber}, MTLD \cite{mccarthy2010mtld}, HDD \cite{mccarthy2007vocd}, Yule's~K \cite{yule1944statistical}, average word length, and a complexity measure. We calculated the complexity as $1 - Flesch Reading Ease$ \cite{flesch1948new}, so higher values indicate harder-to-read, more complex text. 

\subsection{Admission Outcomes and Estimated LLM Usage}

\subsubsection{Difference-in-Differences Analysis}
We employed the difference-in-difference (DiD) framework to characterize whether the magnitude of the SES admissions gap shifted in the post-GPT period, after adjusting for time trends common to both groups. This approach isolates the differential change but does not establish that LLM availability caused the shift because concurrent changes, including post-pandemic recovery and policy responses to Students for Fair Admissions v. Harvard, may have disproportionately affected lower SES applicants during the same period. This approach compares the change in admission rates for lower SES applicants (fee waiver recipients) versus higher SES applicants between the pre-GPT and post-GPT eras. Let $\pi_i = P(\text{Admission}_i = 1 \mid \mathbf{X}_i)$. We estimate:

\begin{equation}\label{did}
\begin{aligned}
\text{logit}(\pi_i)
&= \beta_0 + \beta_1 \text{FeeWaiver}_i + \beta_2 \text{PostGPT}_i \\
&\quad + \beta_3 \left(\text{FeeWaiver}_i \times \text{PostGPT}_i\right)
+ \gamma \mathbf{X}_i + \epsilon_i
\end{aligned}
\end{equation}

where $FeeWaiver_i$ is a binary indicator for lower SES status, $PostGPT_i$ a binary indicator for the post-GPT period, $\mathbf{X}_i$ is a vector of control variables (GPA, test scores, demographics, school type, and honors), $\beta_3$ is the DiD estimator of interest. A negative $\beta_3$ indicates that the admissions gap widened for low-income students in the post-GPT period, after controlling for overall time trends affecting all applicants.

The DiD framework relies on the assumption that, absent widely accessible LLM tools, admission outcomes for lower- and higher SES applicants would have followed parallel trends over time. It also assumes no contemporaneous changes disproportionately affecting one group; to mitigate this risk, models control for academic credentials, demographics, and school characteristics. We assess these assumptions through a series of formal checks, including an event study specification, placebo timing tests, rolling-window estimation, donut-hole analysis, COVID interaction tests, and covariate stability diagnostics (Appendix~\ref{appdx:did_formal_check}). These checks reveal that while pre-treatment parallel trends hold for 2020 and 2021, a significant divergence in 2022, likely attributable to COVID-era test-optional policies, complicates causal interpretation. We therefore interpret the DiD estimates as characterizing shifting associations rather than establishing that LLM availability caused the widening gap, and complement the DiD with stratified and interaction models that directly link estimated LLM usage to admission outcomes within the post-GPT period.

\subsubsection{Stratified Models}
To determine whether LLM usage is associated with different admission outcomes for low-income versus high-income applicants, we estimated separate logistic regression models for each income group in the post-GPT era:

\begin{equation}\label{stratified_effect}
\begin{aligned}
\text{logit}(\pi_i) = \beta_0 + \beta_1\hat{\alpha}_i + \gamma \mathbf{X}_i + \epsilon_i
\end{aligned}
\end{equation}

Models are estimated separately for higher SES applicants and lower SES applicants. Comparing the coefficients across models shows whether the association of LLM usage on admission probability differs by socioeconomic status. A more negative coefficient for lower SES applicants would suggest that LLM usage is more detrimental to their admission chances.

\subsubsection{Interaction Models}
To formally test the differential association of LLM usage across income groups, we estimated an interaction model on the post-GPT subsample:

\begin{equation}
\begin{aligned}
\text{logit}(\pi_i)
&= \beta_0 + \beta_1 \text{FeeWaiver}_i + \beta_2 \hat{\alpha}_i \\
&\quad + \beta_3 \left(\text{FeeWaiver}_i \times \hat{\alpha}_i\right)
+ \gamma \mathbf{X}_i + \epsilon_i
\end{aligned}
\end{equation}

where $\beta_2$ represents the association of LLM usage on admission probability for higher SES students (baseline), $\beta_3$ represents the additional relationship for low-income students, and the total association for lower SES students is $\beta_2 + \beta_3$. A statistically significant $\beta_3$ coefficient provides evidence that the relationship between LLM usage and admission outcomes varies by socioeconomic status.

\subsubsection{Meditation Analysis with Stylometric Features.}
To explore whether observable differences in writing characteristics help explain the differential association between LLM usage and admission outcomes across SES groups, we follow a change-in-coefficient approach \cite{vanderweele2015explanation}. Specifically, we compare the magnitude of the interaction term $\beta_3$ (FeeWaiver $\times$ $\hat{\alpha}$) across two nested models: a baseline model and an augmented model that additionally controls for a vector of 11 stylometric features $\mathbf{S}_i$:

\begin{equation}
\begin{aligned}
\text{logit}(\pi_i) 
&= \beta_0 + \beta_1 \text{FeeWaiver}_i + \beta_2 \hat{\alpha}_i \\
& + \beta_3 (\text{FeeWaiver}_i \times \hat{\alpha}_i) + \boldsymbol{\delta} \mathbf{S}_i + \boldsymbol{\gamma} \mathbf{X}_i + \epsilon_i
\label{eq:augmented}
\end{aligned}
\end{equation}

If the stylometric features partially account for the differential association, $|\beta_3|$ should attenuate when $\mathbf{S}_i$ is included. The percentage reduction in $|\beta_3|$ provides an approximate measure of the share of the differential association attributable to observable linguistic differences between SES groups among LLM users.

We note that because both $\hat{\alpha}$ and $\mathbf{S}_i$ characterize the same essay, they are likely jointly determined, which limits causal interpretation under the sequential ignorability assumption \cite{imai2010general, tingley2014mediation}. The change-in-coefficient approach in logistic regression can also conflate mediation with rescaling effects across nested nonlinear models \cite{karlson2012comparing}. We therefore treat this as an exploratory mechanism analysis and focus on the direction and relative magnitude of pathways rather than precise point estimates.

To complement the change-in-coefficient analysis, we provide Average Causal Mediation Effects (ACME) and Average Direct Effects (ADE) for each stylometric feature, providing a more granular decomposition of how observable linguistic dimensions contribute to the differential association. A negative ACME indicates mediation, in which the feature partially accounts for the differential association between LLM usage and admission outcomes. A positive ACME indicates suppression, in which the feature masks a larger underlying disparity: the observed differential understates the penalty once that dimension is held constant.

\section{Results}
\subsection{Linguistic Convergence Across Socioeconomic and Admissions Groups}

\begin{figure*}[!t]
    \centering
    \begin{subfigure}[t]{0.75\linewidth}
        \centering
        \includegraphics[width=\linewidth]{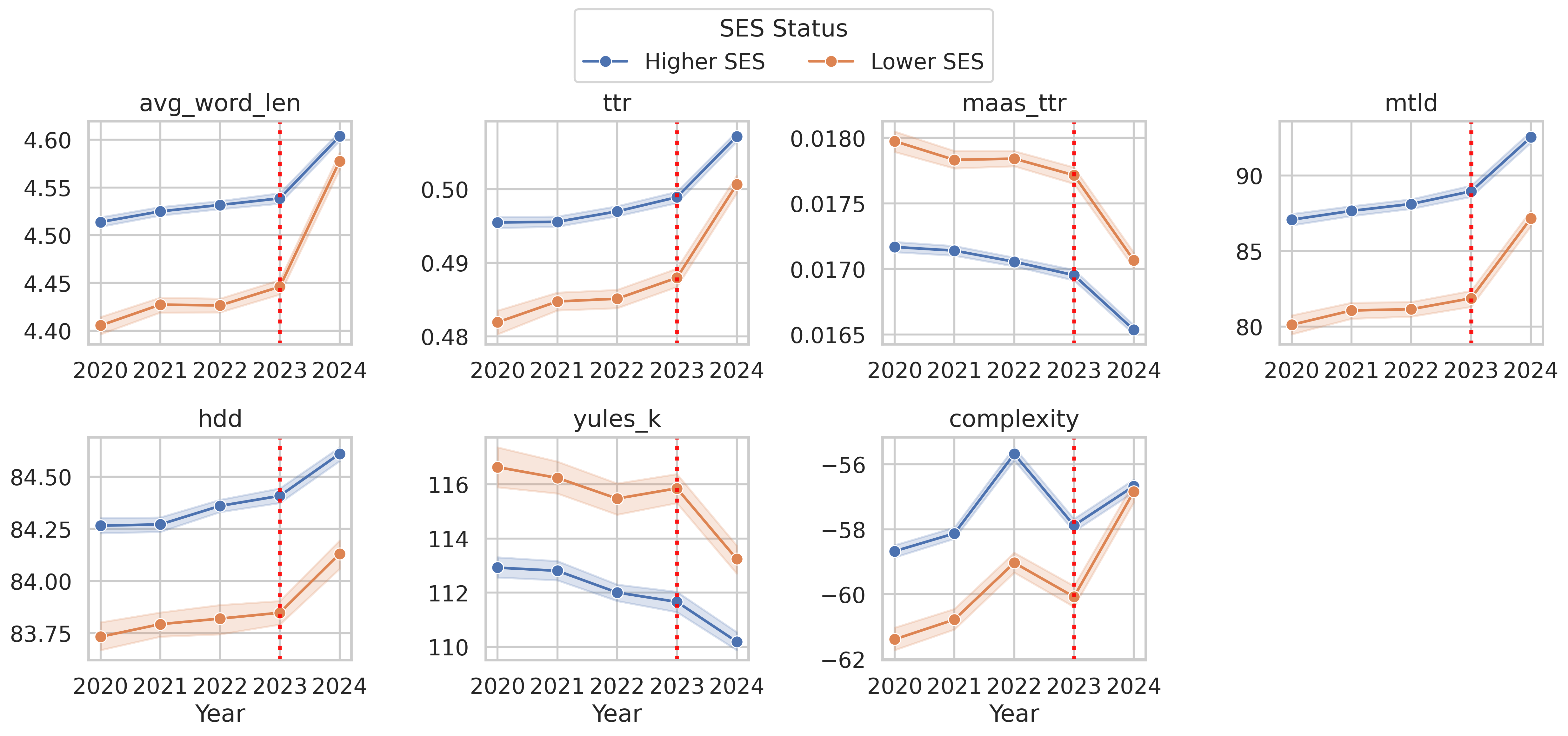}
        \caption{Lexical Diversity and Complexity by Socioeconomic Status (Fee Waiver)}
        \label{fig:ling_change_over_yrs}
    \end{subfigure}

    \vspace{0.6em}

    \begin{subfigure}[t]{0.75\linewidth}
        \centering
        \includegraphics[width=\linewidth]{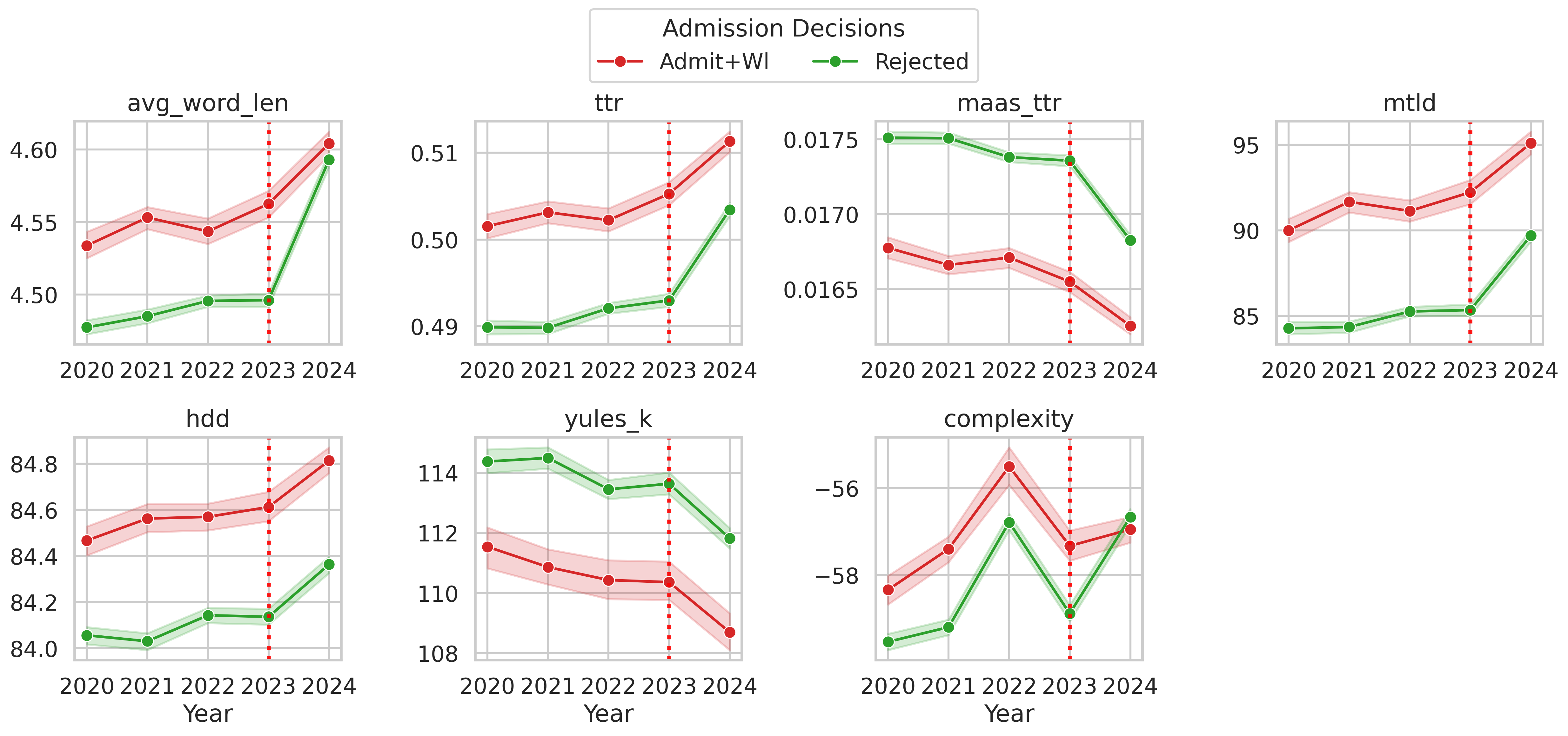}
        \caption{Lexical Diversity and Complexity by Admission Outcome (Admitted/Waitlisted vs. Rejected)}
        \label{fig:ling_change_over_yrs_decisions}
    \end{subfigure}

    \caption{Linguistic changes in college admission essays over time (2020--2024), stratified by socioeconomic status (a) and admission outcome (b). Essays exhibit substantial post-2023 convergence in surface-level linguistic features, with larger changes among lower SES and rejected applicants. Shaded regions indicate 95\% confidence intervals; the vertical dashed line marks ChatGPT's public release (Nov.30, 2022).}
    \label{fig:ling_change_over_yrs_overall}
\end{figure*}

To examine how linguistic features of admission essays have changed over time across socioeconomic and admissions groups (RQ1), Figure~\ref{fig:ling_change_over_yrs_overall} presents the temporal evolution of lexical diversity and complexity measures from the 2019--2020 cycle to the 2023--2024 cycle, stratified by both socioeconomic status and admissions outcome. Panel~\ref{fig:ling_change_over_yrs} reveals that high and lower SES applicants initially exhibited distinct linguistic profiles in 2020, with essays from lower SES background applicants demonstrating lower average word length, higher type-token ratios, and reduced complexity scores. However, these disparities diminished substantially over the subsequent years, with both groups converging toward similar linguistic profiles by 2024. This convergence suggests a homogenization of writing styles across socioeconomic backgrounds, potentially driven by increased access to essay coaching, generative AI tools, or standardized preparation resources. Panel~\ref{fig:ling_change_over_yrs_decisions} demonstrates a parallel pattern when comparing admitted and rejected applicants: while early-cycle essays showed modest linguistic differentiation between the two groups, with admitted students producing slightly more diverse vocabulary and complex sentence structures, these differences largely disappeared by 2024. The shaded regions indicating 95\% confidence intervals show substantial overlap between groups in the later years, suggesting that there is no statistically significant difference between groups. Notably, the convergence accelerated markedly after 2023, coinciding with the widespread adoption of ChatGPT and similar large language models, though we reserve detailed analysis of AI influence for subsequent sections. 


\begin{table}[!t]
    \centering
    \caption{Linguistic Features by SES Status in the Post-GPT Period}
    \label{tab:stylometric_descriptives}
    \resizebox{\columnwidth}{!}{%
    \begin{tabular}{l c c c c}
        \toprule
        \textbf{Feature} & \textbf{Higher SES} & \textbf{Lower SES} & \textbf{Difference} & \textbf{$t$ stat.} \\
         & \textbf{Mean (SD)} & \textbf{Mean (SD)} & \textbf{(\% diff)} &  \\
        \midrule

        \# tokens             & 627.91 (47.32) & 607.69 (74.46) & -20.21 (-3.2\%)*** & -16.84 \\
        \# words              & 621.03 (46.61) & 601.00 (73.39) & -20.04 (-3.2\%)*** & -16.94 \\
        \# types              & 319.84 (33.38) & 306.21 (41.90) & -13.64 (-4.3\%)*** & -17.89 \\
        Avg.\ word length     & 4.71 (0.27) & 4.72 (0.32) & 0.00 (0.1\%)        & 0.56 \\
        Avg.\ sentence length & 19.01 (3.55) & 18.95 (3.74) & -0.06 (-0.3\%)      & -0.73 \\
        TTR                   & 0.52 (0.04) & 0.51 (0.04) & 0.00 (-0.9\%)***   & -5.61 \\
        MAAS\_TTR             & 0.02 (0.00) & 0.02 (0.00) & 0.00 (2.7\%)***     & 10.75 \\
        MTLD                  & 95.78 (21.39) & 90.88 (19.79) & -4.90 (-5.1\%)***   & -11.13 \\
        HDD                   & 84.81 (1.85) & 84.34 (1.88) & -0.47 (-0.6\%)***   & -11.97 \\
        Yules' K              & 108.42 (18.67) & 112.00 (18.84) & 3.58 (3.3\%)***     & 9.07 \\
        Complexity            & -53.07 (9.83) & -52.47 (11.16) & 0.60 (-1.1\%)**     & 2.76 \\
        \midrule
        N                     &   7{,}377     &   3{,}249      &                    & \\
        \bottomrule
    \end{tabular}}
    \begin{tablenotes}[flushleft]
        \footnotesize
        \item *$p < 0.05$, **$p < 0.01$, ***$p < 0.001$
        \item \% diff = percentage difference relative to higher SES mean.
    \end{tablenotes}
\end{table}

Focusing on the post-GPT period specifically, Table~\ref{tab:stylometric_descriptives} provides a detailed comparison of linguistic features between higher SES and lower SES applicants among those with detectable LLM usage ($\hat{\alpha} > 0$) in the 2023--2024 cycle. While Figure~\ref{fig:ling_change_over_yrs_overall} documents overall convergence trends, examining LLM users separately reveals that meaningful differences persist even after accounting for AI adoption. lower SES LLM users produce essays that are 3.2\% shorter and exhibit 4.3\% fewer unique word types, 5.1\% lower MTLD scores (90.88 vs. 95.78, $t = -11.13$, $p < 0.001$), and 3.3\% higher Yule's K values (112.00 vs. 108.42, $t = 9.07$, $p < 0.001$), indicating reduced lexical diversity and greater vocabulary repetition. These patterns suggest that despite widespread LLM adoption contributing to the overall convergence observed in Figure~\ref{fig:ling_change_over_yrs_overall}, lower SES students use these tools differently or less effectively than their higher SES peers.

\subsection{Lower SES and Rejected Applicants Show Steeper LLM Adoption}

\begin{figure}[!htpb]
    \centering

    \begin{subfigure}[t]{0.95\linewidth}
        \centering
        \includegraphics[width=\linewidth]{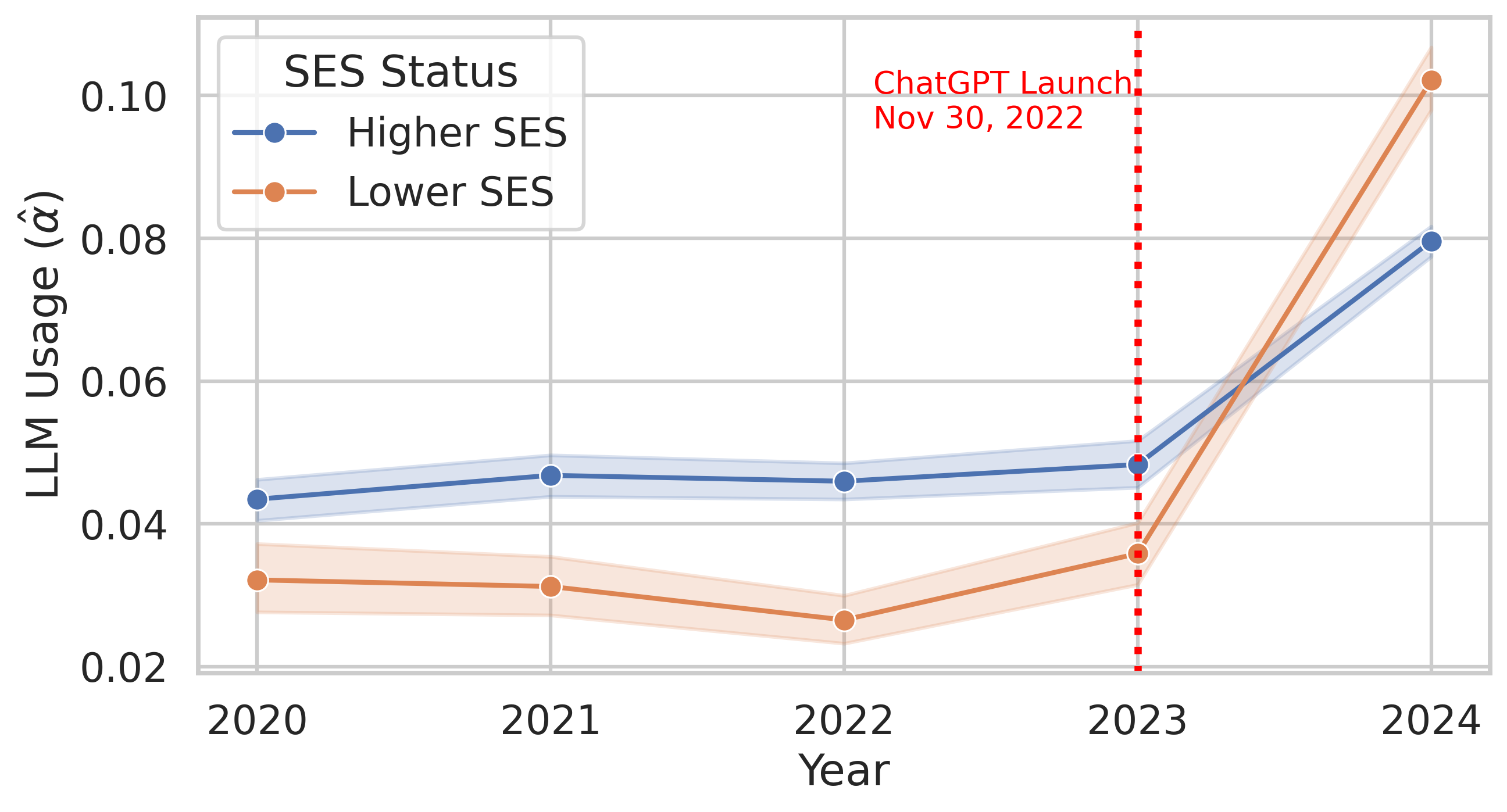}
        \caption{LLM Usage by Socioeconomic Status}
        \label{fig:est_alpha_fee_waiver}
    \end{subfigure}

    \vspace{0.6em}

    \begin{subfigure}[t]{0.95\linewidth}
        \centering
        \includegraphics[width=\linewidth]{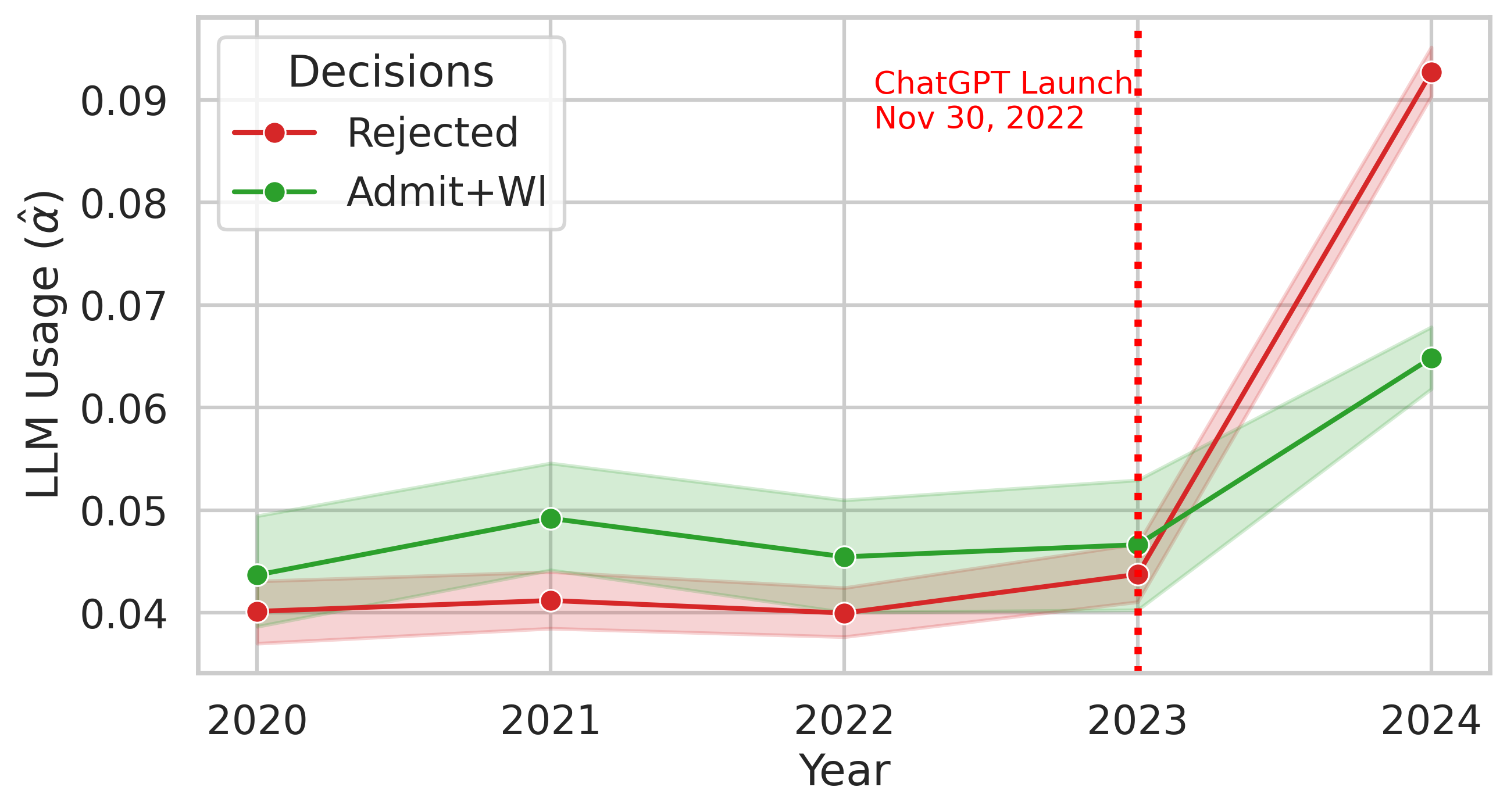}
        \caption{LLM Usage by Admission Outcome (Admitted/Waitlisted vs. Rejected)}
        \label{fig:est_alpha_decisions}
    \end{subfigure}

    \caption{Estimated LLM usage ($\hat{\alpha}$) in admission essays (2020--2024), by socioeconomic status and admission outcome. Usage rises sharply after 2023, with larger increases among lower SES and rejected applicants. Shaded regions denote 95\% confidence intervals; the dashed line marks ChatGPT's public release.}
    \label{fig:est_alpha_both}
\end{figure}

Turning to how LLM-assisted writing adoption has evolved across socioeconomic and admissions groups (RQ2), Figure~\ref{fig:est_alpha_both} presents our estimated fraction of LLM-generated text in admission essays from 2020 to 2024, revealing a sharp shift in writing practices following the public release of ChatGPT in November 2022.Panel \ref{fig:est_alpha_fee_waiver} shows that LLM usage remained relatively flat and low across both lower and higher SES status applicants through 2023, with estimated $\hat{\alpha}$ values hovering around 0.03 to 0.05. However, in 2024, both groups exhibit sharp increases in estimated LLM usage, with lower SES applicants showing a steeper trajectory, reaching approximately 0.10 (10\% LLM-generated content) compared to 0.08 for higher SES applicants. Panel \ref{fig:est_alpha_decisions} demonstrates an even more pronounced divergence when stratifying by admissions outcome: while admitted and waitlisted applicants show modest, parallel increases in LLM usage (reaching $\hat{\alpha} \approx 0.07$ by 2024), rejected applicants exhibit substantially higher adoption rates, with estimated LLM content approaching 0.10 by 2024. 

\subsubsection{LLM Usage Patterns by SES Status}
\begin{figure}[!htbp]
    \centering
    \includegraphics[width=0.45\textwidth]{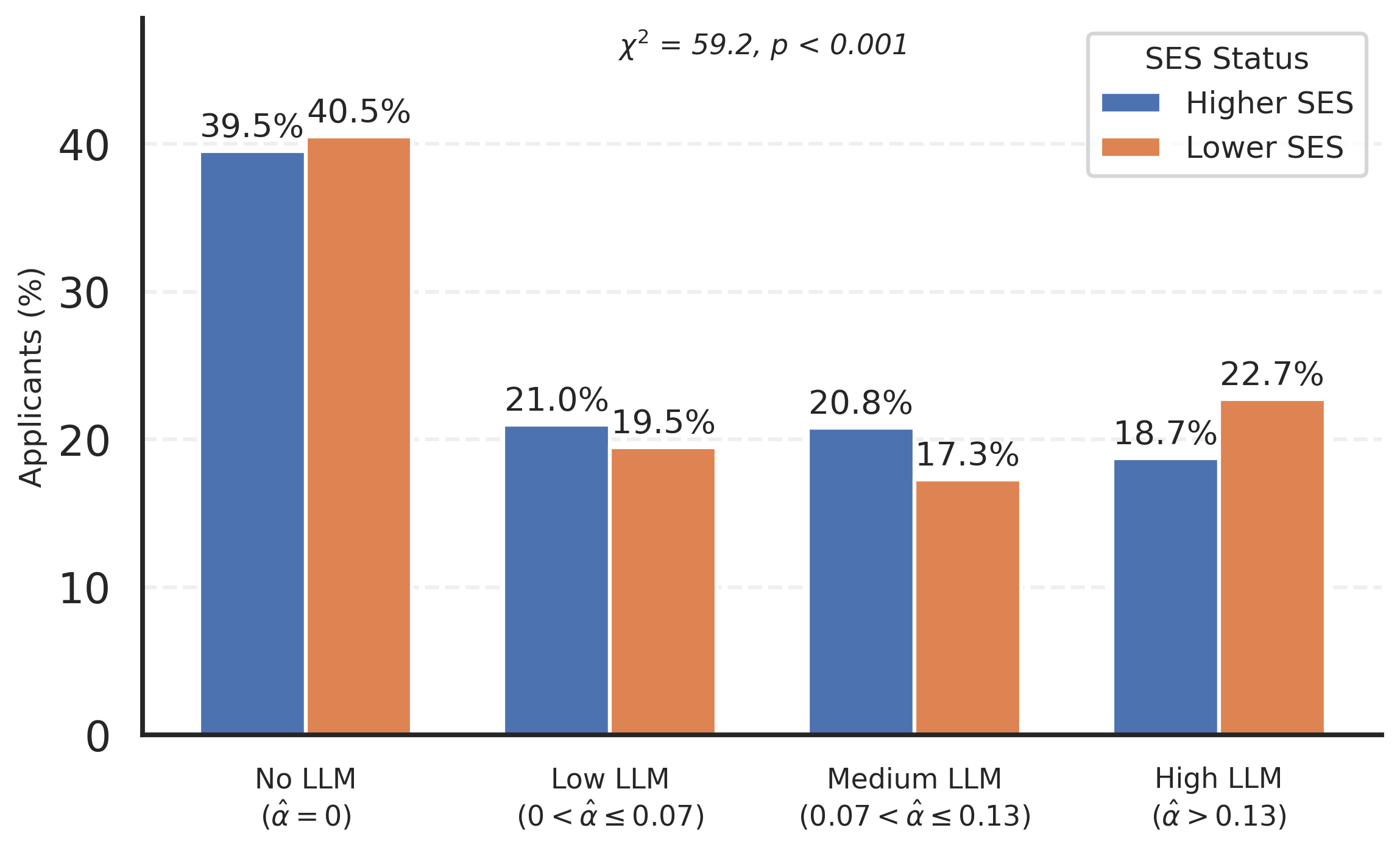}
    \caption{LLM Usage Distribution by SES Status in the Post-GPT Period.}
    \label{fig:llm_usage_dist_by_ses}
\end{figure}

We further examine differences in LLM usage across SES groups in the post-GPT period. Figure~\ref{fig:est_alpha_both} and \ref{fig:llm_usage_dist_by_ses} present LLM usage patterns across socioeconomic groups in the post-GPT subsample. Lower SES applicants exhibited significantly higher mean estimated $\hat{\alpha}$ values ($M = 0.102$, $SD = 0.169$) compared to non-waiver applicants ($M = 0.080$, $SD = 0.119$), $t(17{,}652) = 10.17$, $p < 0.001$, Cohen's $d = 0.15$). This 2.3 percentage point (pp) difference, while modest in magnitude, represents a 28\% higher rate of detectable LLM usage among lower SES students.

Examining categorical usage patterns reveals pronounced SES-based disparities (Figure~\ref{fig:llm_usage_dist_by_ses}). Higher and lower SES applicants showed nearly identical rates of no detectable LLM usage (39.5\% vs. 40.5\%), but their distributions diverge substantially across the usage categories ($\chi^2 = 59.2$, $df = 3$, $p < 0.001$). In the low and medium usage categories, higher SES applicants are slightly overrepresented by 1.5 and 3.5 percentage points, respectively. This pattern reverses sharply at high usage ($\hat{\alpha} > 0.13$), where lower SES applicants are overrepresented by 4 percentage points, or roughly 21\% relative to their higher SES counterparts (22.7\% vs. 18.7\%). These patterns suggest that lower SES students not only adopted LLMs at higher rates overall but were particularly concentrated in the high-intensity usage category, potentially reflecting greater reliance on LLMs for essay composition.

\subsection{Admission Outcomes and Estimated LLM Usage}

\subsubsection{DiD Analysis}

\begin{figure}[ht]
    \centering
    \includegraphics[width=0.35\textwidth]{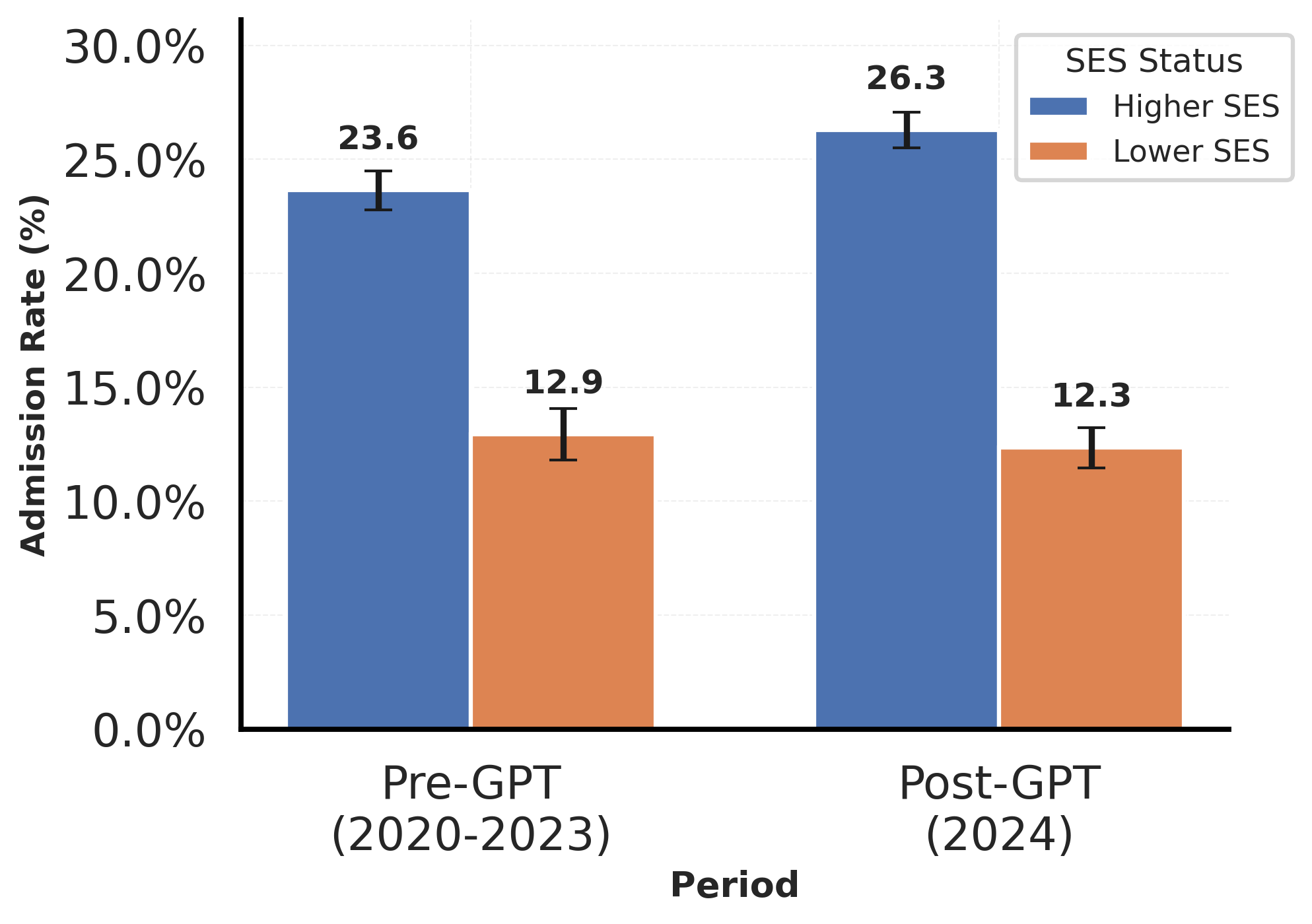}
    \caption{Admission rates by socioeconomic status across pre-GPT and post-GPT eras.}
    \label{fig:did_analysis}
\end{figure}

To examine whether the relationship between essay characteristics and admission outcomes shifted following widespread LLM adoption (RQ3), we first assess overall trends using a difference-in-differences framework. The socioeconomic gap in admissions widened in the post-GPT era. Figure~\ref{fig:did_analysis} and Table~\ref{tab:did_results} show that lower SES applicants had substantially lower admission rates than higher SES applicants prior to ChatGPT’s release (12.9\% vs. 23.6\%), a gap of 10.7 percentage points (OR = 0.572, $p$ < .001). Admission rates increased overall in the post-GPT period (OR = 1.146, $p$ < .001), but this improvement was not evenly distributed across groups. The DiD interaction indicates that the SES gap widened significantly ($\beta$ =  -0.170, OR = 0.844, $p$ = .025), corresponding to an additional 15.6\% reduction in admission odds for lower SES applicants relative to higher SES applicants after LLMs became widely available.

By the 2023–2024 cycle, the admissions gap had expanded to 14.0 percentage points (12.3\% vs. 26.3\%), representing a 31\% increase relative to the pre-GPT baseline. This widening gap reflects diverging trends: admission rates rose for higher SES applicants (+2.7pp) but slightly declined for lower SES applicants (-0.6pp). These patterns persist after controlling for academic credentials, demographics, school type, and leadership/honors participation, indicating that observable applicant characteristics do not fully explain the growing disparity.

\begin{table}[ht]
    \centering
    \caption{Difference-in-Differences: Admission Rates by SES Status}
    \label{tab:did_results}
    \resizebox{\columnwidth}{!}{%
    \begin{tabular}{l c c c c}
        \toprule
        \textbf{Variable} & \textbf{Coef.} & \textbf{OR [95\% CI]} & \textbf{SE} & \textbf{P-value} \\
        \midrule
        \multicolumn{5}{l}{\textit{DiD Components}} \\
        Lower SES ($\beta_1$) & $-0.559$*** & 0.572 [0.506, 0.647] & 0.063 & $<$0.001 \\
        Post-GPT ($\beta_2$) & $0.136$*** & 1.146 [1.074, 1.224] & 0.033 & $<$0.001 \\
        \textbf{Lower SES $\times$ Post-GPT ($\beta_3$)} & $\mathbf{-0.170}$\textbf{*} & \textbf{0.844 [0.727, 0.979]} & \textbf{0.076} & \textbf{0.025} \\
        \addlinespace
        \multicolumn{5}{l}{\textit{Demographics}} \\
        Sex (Ref: Female) & & & & \\
        \hspace{1em}Male & $-0.574$*** & 0.563 [0.530, 0.599] & 0.031 & $<$0.001 \\
        First Generation (Ref: First Gen) \\
        \hspace{1em}Multi Gen & $-0.376$*** & 0.686 [0.624, 0.754] & 0.048 & $<$0.001 \\
        \addlinespace
        \multicolumn{5}{l}{\textit{Academic Performance}} \\
        Cumulative GPA & $0.364$*** & 1.440 [1.381, 1.500] & 0.021 & $<$0.001 \\
        SAT Reading/Writing & $0.865$*** & 2.375 [1.791, 3.149] & 0.144 & $<$0.001 \\
        SAT Math & $-0.486$** & 0.615 [0.463, 0.817] & 0.145 & 0.001 \\
        ACT Composite & $-1.339$*** & 0.262 [0.165, 0.416] & 0.236 & $<$0.001 \\
        ACT Math & $1.622$*** & 5.067 [3.193, 8.030] & 0.235 & $<$0.001 \\
        Leadership/Honors & $0.813$*** & 2.255 [2.056, 2.472] & 0.047 & $<$0.001 \\
        \bottomrule
    \end{tabular}}
    \begin{tablenotes}[flushleft]
        \footnotesize
        \item \textit{Note:} $N = 30{,}488$. Pseudo $R^2 = 0.096$. OR = Odds Ratio; CI = Confidence Interval; SE = Standard Error. *$p < 0.05$, **$p < 0.01$, ***$p < 0.001$.
        \item All continuous academic variables are standardized (mean $= 0$, SD $= 1$).
        \item $\beta_3$ is the DiD estimator. The negative coefficient indicates the SES admissions gap widened in the post-GPT era.
    \end{tablenotes}
\end{table}

\subsubsection{Stratified Models: Differential Associations of LLM Usage by SES Status}

To examine whether the association between estimated LLM usage and admission outcomes differs by SES, we estimated separate logistic regression models for lower and higher SES applicants in the post-GPT era, controlling for GPA, test scores, demographics, school type, and leadership participation. The results show that LLM usage is negatively associated with admission probability for all applicants, but the penalty is substantially larger for lower SES students.

For higher SES applicants, estimated LLM usage demonstrates a significant negative association with admission probability ($\beta = -0.96$, $p < 0.001$; OR = 0.38), corresponding to a 62\% reduction in admission odds per unit increase in $\hat{\alpha}$. While substantial, this penalty is considerably smaller than that experienced by lower SES applicants.

For lower SES applicants, the penalty associated with LLM usage is nearly double in magnitude ($\beta = -1.78$, $p < 0.001$; OR = 0.17), corresponding to an 83\% reduction in admission odds per unit increase in $\hat{\alpha}$. The difference in coefficients ($\Delta\beta = -0.82$) indicates that the penalty is 1.85 times larger for lower SES applicants. This differential association persists even after controlling for GPA, standardized test scores, demographic characteristics, school type, and leadership participation.

These differential associations help explain the widening socioeconomic gap observed in the post-GPT era. Lower SES students not only adopted LLMs at higher rates (22.7\% vs. 18.7\% in the high-use category) but also experienced larger negative associations between estimated LLM usage and admission outcomes. This combination of higher adoption and more severe consequences creates a compounding disadvantage that contributes to the 3.3pp widening of the admissions gap. Full results are reported in Appendix \ref{appdx:stratified associations} Table~\ref{tab:stratified_effects}.

\subsubsection{Interaction Model for Differential Relationships of LLM Usage by SES Status}

\begin{table}[ht]
    \centering
    \caption{Interaction Model: Differential Associations of LLM Usage by SES Status}
    \label{tab:interaction_model}
    \resizebox{\columnwidth}{!}{%
    \begin{tabular}{l c c c c}
        \toprule
        \textbf{Variable} & \textbf{Coefficient} & \textbf{OR [95\% CI]} & \textbf{SE} & \textbf{P-value} \\
        \midrule
        \multicolumn{5}{l}{\textit{Key Variables}} \\
        Lower SES & $-0.625$*** & 0.536 [0.474, 0.606] & 0.063 & $<$0.001 \\
        $\hat{\alpha}$ (higher SES baseline, $\beta_2$) & $-0.922$*** & 0.398 [0.264, 0.600] & 0.209 & $<$0.001 \\
        \textbf{Lower SES $\times$ $\hat{\alpha}$ (Interaction $\beta_3$)} &\textbf{ $-1.028$*} & \textbf{0.358 [0.147, 0.868]} & \textbf{0.452} &\textbf{ 0.023} \\
        \hspace{1em} \textit{Total $\hat{\alpha}$ Association ($\beta_2$+$\beta_3$)} & $-1.949$*** & 0.142 [0.065, 0.313] & 0.402 & $<$0.001 \\
        \addlinespace
        \multicolumn{5}{l}{\textit{Demographics}} \\
        Sex (Ref: Female) & & & & \\
        \hspace{1em} Male & $-0.489$*** & 0.613 [0.565, 0.665] & 0.041 & $<$0.001 \\
        First Generation (Ref: First Gen) & & & & \\
        \hspace{1em} Multi Gen & $-0.455$*** & 0.635 [0.560, 0.720] & 0.064 & $<$0.001 \\
        \addlinespace
        \multicolumn{5}{l}{\textit{School Type (Ref: Home)}} \\
        \hspace{1em} Private & 0.384 & 1.467 [0.559, 3.853] & 0.492 & 0.436 \\
        \hspace{1em} Public & 0.735 & 2.085 [0.797, 5.454] & 0.491 & 0.134 \\
        \hspace{1em} Unknown & $-0.103$ & 0.903 [0.291, 2.801] & 0.578 & 0.859 \\
        \addlinespace
        \multicolumn{5}{l}{\textit{Academic Performance}} \\
        Cumulative GPA & 1.048*** & 2.852 [2.459, 3.308] & 0.076 & $<$0.001 \\
        SAT Reading/Writing & 0.002*** & 1.002 [1.001, 1.003] & 0.001 & $<$0.001 \\
        SAT Math & $-0.001$ & 0.999 [0.998, 1.000] & 0.001 & 0.061 \\
        ACT Composite Score & $-0.126$*** & 0.882 [0.838, 0.928] & 0.026 & $<$0.001 \\
        ACT Math Score & 0.154*** & 1.167 [1.110, 1.226] & 0.026 & $<$0.001 \\
        Leadership/Honors & 0.785*** & 2.192 [1.931, 2.489] & 0.065 & $<$0.001 \\
        \addlinespace
        Intercept & $-3.331$*** & 0.036 [0.013, 0.096] & 0.502 & $<$0.001 \\
        \bottomrule
    \end{tabular}}
    \begin{tablenotes}[flushleft]
        \footnotesize
        \item \textit{Note:} $N = 17{,}654$. Pseudo $R^2 = 0.112$. *$p < 0.05$, **$p < 0.01$, ***$p < 0.001$.
        \item Total $\hat{\alpha}$ association (lower SES) = $\beta_2 + \beta_3$ represents the combined association of LLM usage for lower SES students.
        \item The interaction term ($\beta_3$) tests whether the association of LLM usage differs by SES. Sample restricted to post-GPT era (2023--2024 admissions cycle).
    \end{tablenotes}
\end{table}

Table~\ref{tab:interaction_model} presents results from a logistic regression model that formally tests whether the association between LLM usage and admission outcomes differs by socioeconomic status by estimating an interaction between SES status and LLM usage ($\hat{\alpha}$).

We find that the negative association between LLM usage and admission probability is significantly stronger for lower SES applicants. For higher SES students (baseline group), estimated LLM usage is associated with lower admission odds ($\beta_2 = -0.92$, OR = 0.40, $p < 0.001$), corresponding to an approximate 60\% reduction in admission odds per unit increase in $\hat{\alpha}$. The interaction term shows that this penalty is significantly larger for lower SES applicants ($\beta_3 = -1.03$, OR = 0.36, $p = 0.023$), providing evidence that the relationship between LLM usage and admission outcomes differs by socioeconomic status. Combining the baseline and interaction terms yields the total association for lower SES students ($\beta_2 + \beta_3 = -1.95$), corresponding to an 85.8\% reduction in admission odds per unit increase in $\hat{\alpha}$, more than twice the penalty observed for higher SES students (ratio = 2.12).

To illustrate the practical magnitude of these relationships, consider two applicants with identical observable credentials (same GPA, test scores, demographics, etc.) who both produce essays at the high-use threshold ($\hat{\alpha} = 0.13$). A higher SES applicant would experience a predicted reduction in admission odds of approximately 11\% relative to a non-user with the same credentials. A lower SES applicant with the same estimated LLM usage would experience a reduction of approximately 22\%. At $\hat{\alpha} = 0.20$, roughly the upper range of typical LLM users in our sample, the gap widens further: approximately 17\% for higher SES and 33\% for lower SES applicants.

These findings help explain the widening socioeconomic gap documented in our difference-in-differences analysis (Table~\ref{tab:did_results}). The 3.3pp post-GPT widening reflects the joint contribution of two compounding factors: differential adoption, with lower SES students approximately 21\% more likely to fall into the high-intensity LLM usage category, and differential penalties, with lower SES students facing approximately twice the admission odds reduction per unit of estimated LLM usage relative to their higher SES peers.
\subsubsection{Mediation Analysis with Stylometric Features}

\begin{table}[t]
\centering
\caption{Mediation Analysis Results with Average Causal Mediation Effect (ACME) and Average Direct Effect (ADE). Negative ACME indicates mediation (feature partially explains the penalty); positive ACME indicates suppression (feature masks a larger penalty).}
\label{tab:mediation}
\small
\begin{tabular}{lrrr}
\toprule
\textbf{Feature} & \textbf{ACME} & \textbf{ADE} & \textbf{\% Mediated}  \\
\midrule
\# tokens        & $-0.034$*** & $-0.116$*** & 25.1\%     \\
\# words         & $-0.027$*** & $-0.121$*** & 20.3\%     \\
\# types         &  $0.040$*** & $-0.162$*** & $-27.9$\%  \\
Avg. word length   &  $0.200$*** & $-0.194$*** & $-149.5$\% \\
Avg. sentence length   &  $0.001$    & $-0.128$*** & $-1.2$\%   \\
Complexity       &  $0.127$*** & $-0.179$*** & $-94.2$\%  \\
Maas TTR        &  $0.072$*** & $-0.166$*** & $-52.8$\%  \\
TTR              &  $0.063$*** & $-0.162$*** & $-46.8$\%  \\
MTLD             &  $0.052$*** & $-0.158$*** & $-39.0$\%  \\
HDD              &  $0.019$*** & $-0.144$*** & $-13.9$\%  \\
Yules' k         &  $0.012$*** & $-0.137$*** & $-8.9$\%   \\
\bottomrule
\multicolumn{4}{l}{\footnotesize Stat. significance: *** $p < 0.001$ based on 1{,}000 bootstrap simulations.} \\
\end{tabular}
\end{table}

Table~\ref{tab:mediation} presents results from mediation analyses for each of the 11 stylometric features, treating estimated LLM usage ($\hat{\alpha}$) as the treatment and admission as the outcome. Two features partially mediate the differential SES penalty:  essay length measured in tokens (ACME $= -0.034$, $p < 0.001$; proportion mediated $= 25.1\%$) and word count (ACME $= -0.027$, $p < 0.001$; $20.3\%$). These results suggest that differences in essay length between SES groups among  LLM users account for approximately one fifth to one fourth of the differential admission penalty.

The remaining significant features operate as suppressors rather than mediators, with positive ACME estimates indicating that controlling for these dimensions reveals a larger, not smaller, SES penalty. Unlike mediators that explain a portion of the total effect, suppressors work in the opposite direction: the indirect pathway through these features partially offsets the overall penalty, such that the observed total effect understates the underlying disadvantage once these compensating dimensions are held constant. The strongest suppressors are average word length (ACME $= 0.200$, $p < 0.001$) and syntactic complexity (ACME $= 0.127$, $p < 0.001$). This pattern suggests that on dimensions of lexical sophistication and complexity, lower SES students who use LLMs produce essays that are comparable to or exceed those of higher SES peers, partially offsetting the overall penalty. When these compensating features are held constant, the underlying disadvantage is larger than the raw interaction coefficient implies. Lastly, average sentence length showed no significant indirect pathway (ACME $= 0.001$, $p = 0.724$). Full results are described in Appendix \ref{appdx:mediation}.

The persistence of a significant interaction after controlling for linguistic features strengthens the interpretation that the disparities reflect differences in how LLM use is evaluated, rather than pre-existing writing differences that correlate with SES.
Addressing the inequitable impacts of LLM tools will require more than helping students produce similar surface-level linguistic prose; the differential penalty likely arises from deeper judgments about authenticity, quality, and authorship in essay evaluation.

\section{Discussion}
This study examined generative AI through the lens of the digital divide, which distinguishes between inequalities in access, use, and outcomes. Using longitudinal admissions data spanning the pre- and post-GPT era, we investigated three questions: how writing changed over time (RQ1), how LLM adoption differed by socioeconomic status (RQ2), and whether the relationship between essays and admission outcomes shifted (RQ3). Our findings suggest a transition from a traditional divide in access to writing support toward a more complex divide in how AI assistance translates into evaluative outcomes.

\paragraph{Linguistic convergence across groups.}
Admission essays underwent substantial linguistic convergence after 2023, with surface-level features narrowing across socioeconomic and admissions groups (Figure \ref{fig:ling_change_over_yrs}). This convergence was driven primarily by larger changes among lower SES and rejected applicants, consistent with evidence that LLM-assisted text homogenizes toward common distributional patterns \cite{doshi2024generative, anderson2024homogenization, moon2024homogenizing} and dominant cultural norms \cite{agarwal2025ai, alvero2025digital}. In admissions, where essays are intended to surface individual voice \cite{beck2023makes, aukerman2018student}, this flattening erodes the signal value of personal statements in holistic review.

\paragraph{Differential LLM adoption by SES groups.}
Estimated LLM use rose sharply in 2024 across all applicants but increased disproportionately among lower SES students. These lower SES applicants were 28\% more likely to fall into the high-intensity usage category ($\hat{\alpha} > 0.13$), consistent with an access-substitution mechanism: for students with limited access to private tutors or essay coaches, LLMs may function as a substitute for scarce support \cite{early2011making, warren2013rhetoric, huang2024translating}. This aligns with experimental evidence that generative AI disproportionately benefits lower-skilled workers \cite{noy2023experimental, brynjolfsson2025generative} and with survey data showing rising student AI adoption \cite{freeman2025student}. Higher adoption among lower SES applicants represents a rational response to resource constraints, a form of technological leveling the digital divide literature has long anticipated \cite{warschauer2004technology}.

\paragraph{Differential associations between LLM use and admission outcomes.}
Yet this leveling in access did not translate into leveling in outcomes. Despite higher adoption, LLM usage was associated with significantly larger admission penalties for lower SES applicants. The total association for lower SES students ($\beta_2 + \beta_3 = -1.95$, OR = 0.14) was more than twice the penalty for higher SES students ($\beta_2 = -0.92$, OR = 0.40), even after controlling for academic credentials. Incorporating stylometric features attenuated the interaction by only 7.1\%, indicating that the disparity operates through mechanisms our lexical and syntactic measures do not capture. We note that the post-GPT period also follows the recovery from COVID-19-era disruptions to schooling and college preparation, which disproportionately affected lower SES students \cite{engzell2021learning, aucejo2020impact}. While our controls for academic credentials and demographics help account for compositional shifts, we cannot fully disentangle the effects of pandemic recovery from those of LLM adoption, and the widening gap we observe may partly reflect compounding disadvantages that preceded widespread AI use.

Several mechanisms may explain this differential. Lower SES students, who tend to use more concrete prompting strategies \cite{bassignana2025ai, yu2024whose}, may produce LLM-assisted text that retains more detectable markers of AI involvement compared to higher SES students who integrate suggestions more seamlessly. This is consistent with evidence that effective LLM use requires AI literacy that is itself stratified by educational experience \cite{daepp2025emerging}. Additionally, admissions readers may interpret polished writing with greater suspicion when it appears inconsistent with other aspects of a lower SES applicant's profile \cite{huang2024translating, stevens2009creating, bastedo2022contextualizing}. LLMs may also reproduce linguistic norms associated with privilege \cite{alvero2024large, lee2025poor, hofmann2024ai}, generating text that reads as performatively sophisticated without the narrative specificity readers associate with genuine voice \cite{hwang202580, kadoma2024generative}.

The combination of higher adoption and larger penalties produces a compounding disadvantage that mirrors concerns in the broader digital divide literature: access alone is insufficient when returns to use are unequal \cite{hargittai2002second, daepp2025emerging, sabnis2022large}. This dynamic also echoes findings that AI tools can simultaneously boost individual output while producing unequal downstream consequences \cite{kleinberg2021algorithmic, kusumegi2025scientific}.

These findings carry implications for how essay-based signals are interpreted as AI-assisted writing becomes the norm. The convergence documented in Figure~\ref{fig:ling_change_over_yrs} suggests that surface-level linguistic features are becoming less informative as differentiators. Institutions face difficult choices: whether to discount essays, require disclosure of AI assistance, or redesign prompts and rubrics to foreground dimensions like narrative specificity, experiential detail, and reflective depth that LLMs replicate less convincingly \cite{cho2025academic, nghiem2025rich}. More broadly, our results suggest that the evaluative infrastructure of selective admissions, designed for an era in which writing reliably signaled individual effort, may be poorly suited for an era in which text production is increasingly AI-mediated. The challenge is not simply that students use LLMs, but that the consequences of doing so are distributed unequally \cite{viberg2024advancing, yan2024practical}.

\section{Limitations and Future Work}
Our findings should be interpreted in light of several important limitations that point toward productive directions for future research. First, the distribution-based LLM detector we employed, while validated and consistent with prior work, provides only an estimate rather than definitive proof of individual usage. Future work could benefit from ground-truth data collection such as experimental studies to validate detection methods and better understand the gap between estimated and actual usage patterns. Second, the observational nature of our analysis means we document associations between estimated LLM usage and admission outcomes rather than establishing causal relationships. Experimental studies, such as randomized audits where identical essays are submitted with varying levels of AI assistance, or natural experiments leveraging policy changes in AI tool availability, would help determine whether LLM usage directly affects admission decisions or whether our findings reflect unobserved confounding factors. Third, the generalizability of our results is also constrained by our single-institution focus on a highly selective engineering school. Multi-institutional studies encompassing diverse selectivity levels, disciplinary contexts, and national settings would reveal whether the patterns we observe represent a generalizable phenomenon or are specific to particular institutional and cultural contexts. Finally, the timing of our study introduces an unavoidable confound. ChatGPT's proliferation coincided with the Supreme Court's \textit{Students for Fair Admissions v. Harvard decision} in June 2023, which fundamentally altered race-conscious admissions policies. Disentangling these concurrent shifts would require longitudinal analysis spanning multiple admissions cycles, ideally with institutional variation in how policies responded to the Court's decision.

\section{Conclusion}
In this study, we examined 81,663 college applications and found that lower SES students adopted LLMs at higher rates yet experienced more than twice the admission penalty per unit of estimated usage compared to their higher SES peers, even as LLM use was associated with improved surface-level writing features across groups. This differential persists after controlling for academic credentials and stylometric features, suggesting that it operates through mechanisms beyond text characteristics alone. Although our observational design cannot establish causality, the consistent pattern across adoption, linguistic change, and evaluative outcomes raises concern that widespread LLM use may reshape how essay-based signals function in admissions. Viewed through the digital divide framework, our results suggest a shift from inequalities in access to inequalities in returns, underscoring the need for institutions to reassess how essay-based evidence is interpreted as AI-assisted writing becomes common. Future research should combine experimental, qualitative, and multi-institutional approaches to identify how AI tools interact with existing systems of educational stratification and to inform more equitable evaluation practices.

\section{Acknowledgments}
This research was supported in part by [blinded for review].

\bibliographystyle{ACM-Reference-Format}
\bibliography{ref}


\begin{thebibliography}{61}


\ifx \showCODEN    \undefined \def \showCODEN     #1{\unskip}     \fi
\ifx \showISBNx    \undefined \def \showISBNx     #1{\unskip}     \fi
\ifx \showISBNxiii \undefined \def \showISBNxiii  #1{\unskip}     \fi
\ifx \showISSN     \undefined \def \showISSN      #1{\unskip}     \fi
\ifx \showLCCN     \undefined \def \showLCCN      #1{\unskip}     \fi
\ifx \shownote     \undefined \def \shownote      #1{#1}          \fi
\ifx \showarticletitle \undefined \def \showarticletitle #1{#1}   \fi
\ifx \showURL      \undefined \def \showURL       {\relax}        \fi
\providecommand\bibfield[2]{#2}
\providecommand\bibinfo[2]{#2}
\providecommand\natexlab[1]{#1}
\providecommand\showeprint[2][]{arXiv:#2}

\bibitem[Agarwal et~al\mbox{.}(2025)]%
        {agarwal2025ai}
\bibfield{author}{\bibinfo{person}{Dhruv Agarwal}, \bibinfo{person}{Mor Naaman}, {and} \bibinfo{person}{Aditya Vashistha}.} \bibinfo{year}{2025}\natexlab{}.
\newblock \showarticletitle{AI suggestions homogenize writing toward western styles and diminish cultural nuances}. In \bibinfo{booktitle}{\emph{Proceedings of the 2025 CHI Conference on Human Factors in Computing Systems}}. \bibinfo{pages}{1--21}.
\newblock


\bibitem[Alvero et~al\mbox{.}(2021)]%
        {alvero2021essay}
\bibfield{author}{\bibinfo{person}{AJ Alvero}, \bibinfo{person}{Sonia Giebel}, \bibinfo{person}{Ben Gebre-Medhin}, \bibinfo{person}{Anthony~Lising Antonio}, \bibinfo{person}{Mitchell~L Stevens}, {and} \bibinfo{person}{Benjamin~W Domingue}.} \bibinfo{year}{2021}\natexlab{}.
\newblock \showarticletitle{Essay content and style are strongly related to household income and SAT scores: Evidence from 60,000 undergraduate applications}.
\newblock \bibinfo{journal}{\emph{Science advances}} \bibinfo{volume}{7}, \bibinfo{number}{42} (\bibinfo{year}{2021}), \bibinfo{pages}{eabi9031}.
\newblock


\bibitem[Alvero et~al\mbox{.}(2024)]%
        {alvero2024large}
\bibfield{author}{\bibinfo{person}{AJ Alvero}, \bibinfo{person}{Jinsook Lee}, \bibinfo{person}{Alejandra Regla-Vargas}, \bibinfo{person}{Ren{\'e}~F Kizilcec}, \bibinfo{person}{Thorsten Joachims}, {and} \bibinfo{person}{Anthony~Lising Antonio}.} \bibinfo{year}{2024}\natexlab{}.
\newblock \showarticletitle{Large language models, social demography, and hegemony: comparing authorship in human and synthetic text}.
\newblock \bibinfo{journal}{\emph{Journal of Big Data}} \bibinfo{volume}{11}, \bibinfo{number}{1} (\bibinfo{year}{2024}), \bibinfo{pages}{138}.
\newblock


\bibitem[Alvero et~al\mbox{.}(2025)]%
        {alvero2025digital}
\bibfield{author}{\bibinfo{person}{AJ Alvero}, \bibinfo{person}{Quill Sedlacek}, \bibinfo{person}{Monique Le{\'o}n}, {and} \bibinfo{person}{Cesar Pe{\~n}a}.} \bibinfo{year}{2025}\natexlab{}.
\newblock \showarticletitle{Digital accents, homogeneity-by-design, and the evolving social science of written language}.
\newblock \bibinfo{journal}{\emph{Annual Review of Applied Linguistics}} (\bibinfo{year}{2025}), \bibinfo{pages}{1--19}.
\newblock


\bibitem[Anderson et~al\mbox{.}(2024)]%
        {anderson2024homogenization}
\bibfield{author}{\bibinfo{person}{Barrett~R Anderson}, \bibinfo{person}{Jash~Hemant Shah}, {and} \bibinfo{person}{Max Kreminski}.} \bibinfo{year}{2024}\natexlab{}.
\newblock \showarticletitle{Homogenization effects of large language models on human creative ideation}. In \bibinfo{booktitle}{\emph{Proceedings of the 16th conference on creativity \& cognition}}. \bibinfo{pages}{413--425}.
\newblock


\bibitem[Aucejo et~al\mbox{.}(2020)]%
        {aucejo2020impact}
\bibfield{author}{\bibinfo{person}{Esteban~M Aucejo}, \bibinfo{person}{Jacob French}, \bibinfo{person}{Maria Paola~Ugalde Araya}, {and} \bibinfo{person}{Basit Zafar}.} \bibinfo{year}{2020}\natexlab{}.
\newblock \showarticletitle{The impact of COVID-19 on student experiences and expectations: Evidence from a survey}.
\newblock \bibinfo{journal}{\emph{Journal of public economics}}  \bibinfo{volume}{191} (\bibinfo{year}{2020}), \bibinfo{pages}{104271}.
\newblock


\bibitem[Aukerman and Beach(2018)]%
        {aukerman2018student}
\bibfield{author}{\bibinfo{person}{Maren Aukerman} {and} \bibinfo{person}{Richard Beach}.} \bibinfo{year}{2018}\natexlab{}.
\newblock \showarticletitle{Student conceptualizations of task, audience, and self in writing college admissions essays}.
\newblock \bibinfo{journal}{\emph{Journal of Adolescent \& Adult Literacy}} \bibinfo{volume}{62}, \bibinfo{number}{3} (\bibinfo{year}{2018}), \bibinfo{pages}{319--327}.
\newblock


\bibitem[Bassignana et~al\mbox{.}(2025)]%
        {bassignana2025ai}
\bibfield{author}{\bibinfo{person}{Elisa Bassignana}, \bibinfo{person}{Amanda~Cercas Curry}, {and} \bibinfo{person}{Dirk Hovy}.} \bibinfo{year}{2025}\natexlab{}.
\newblock \showarticletitle{The AI gap: How socioeconomic status affects language technology interactions}. In \bibinfo{booktitle}{\emph{Proceedings of the 63rd Annual Meeting of the Association for Computational Linguistics (Volume 1: Long Papers)}}. \bibinfo{pages}{18647--18664}.
\newblock


\bibitem[Bastedo et~al\mbox{.}(2022)]%
        {bastedo2022contextualizing}
\bibfield{author}{\bibinfo{person}{Michael~N Bastedo}, \bibinfo{person}{Kristen~M Glasener}, \bibinfo{person}{KC Deane}, {and} \bibinfo{person}{Nicholas~A Bowman}.} \bibinfo{year}{2022}\natexlab{}.
\newblock \showarticletitle{Contextualizing the SAT: Experimental evidence on college admission recommendations for low-SES applicants}.
\newblock \bibinfo{journal}{\emph{Educational Policy}} \bibinfo{volume}{36}, \bibinfo{number}{2} (\bibinfo{year}{2022}), \bibinfo{pages}{282--311}.
\newblock


\bibitem[Beck and Godley(2023)]%
        {beck2023makes}
\bibfield{author}{\bibinfo{person}{Sarah~W Beck} {and} \bibinfo{person}{Amanda~J Godley}.} \bibinfo{year}{2023}\natexlab{}.
\newblock \showarticletitle{“What Makes You, You”: The Discursive Construction of the Self in US College Application Essays}.
\newblock \bibinfo{journal}{\emph{American Journal of Education}} \bibinfo{volume}{129}, \bibinfo{number}{4} (\bibinfo{year}{2023}), \bibinfo{pages}{539--564}.
\newblock


\bibitem[Brynjolfsson et~al\mbox{.}(2025)]%
        {brynjolfsson2025generative}
\bibfield{author}{\bibinfo{person}{Erik Brynjolfsson}, \bibinfo{person}{Danielle Li}, {and} \bibinfo{person}{Lindsey Raymond}.} \bibinfo{year}{2025}\natexlab{}.
\newblock \showarticletitle{Generative {AI} at Work}.
\newblock \bibinfo{journal}{\emph{The Quarterly Journal of Economics}} \bibinfo{volume}{140}, \bibinfo{number}{2} (\bibinfo{year}{2025}), \bibinfo{pages}{889--942}.
\newblock


\bibitem[Cho-Baker et~al\mbox{.}(2025)]%
        {cho2025academic}
\bibfield{author}{\bibinfo{person}{Sugene Cho-Baker}, \bibinfo{person}{Brent Bridgeman}, \bibinfo{person}{Guangming Ling}, \bibinfo{person}{Michael Flor}, {and} \bibinfo{person}{Vinetha Belur}.} \bibinfo{year}{2025}\natexlab{}.
\newblock \showarticletitle{Academic Writing Skills in College Admissions Essays: Exploring Their Implications for Admissions Decisions and First-Semester Grade Point Average}.
\newblock \bibinfo{journal}{\emph{Educational Researcher}} (\bibinfo{year}{2025}), \bibinfo{pages}{0013189X251324189}.
\newblock


\bibitem[Daepp and Counts(2025)]%
        {daepp2025emerging}
\bibfield{author}{\bibinfo{person}{Madeleine I.~G. Daepp} {and} \bibinfo{person}{Scott Counts}.} \bibinfo{year}{2025}\natexlab{}.
\newblock \showarticletitle{The Emerging Generative Artificial Intelligence Divide in the United States}. In \bibinfo{booktitle}{\emph{Proceedings of the International AAAI Conference on Web and Social Media}}, Vol.~\bibinfo{volume}{19}. \bibinfo{pages}{443--456}.
\newblock


\bibitem[Deng et~al\mbox{.}(2025)]%
        {deng2025does}
\bibfield{author}{\bibinfo{person}{Ruiqi Deng}, \bibinfo{person}{Maoli Jiang}, \bibinfo{person}{Xinlu Yu}, \bibinfo{person}{Yuyan Lu}, {and} \bibinfo{person}{Shasha Liu}.} \bibinfo{year}{2025}\natexlab{}.
\newblock \showarticletitle{Does ChatGPT enhance student learning? A systematic review and meta-analysis of experimental studies}.
\newblock \bibinfo{journal}{\emph{Computers \& Education}}  \bibinfo{volume}{227} (\bibinfo{year}{2025}), \bibinfo{pages}{105224}.
\newblock


\bibitem[Doshi and Hauser(2024)]%
        {doshi2024generative}
\bibfield{author}{\bibinfo{person}{Anil~R Doshi} {and} \bibinfo{person}{Oliver~P Hauser}.} \bibinfo{year}{2024}\natexlab{}.
\newblock \showarticletitle{Generative AI enhances individual creativity but reduces the collective diversity of novel content}.
\newblock \bibinfo{journal}{\emph{Science advances}} \bibinfo{volume}{10}, \bibinfo{number}{28} (\bibinfo{year}{2024}), \bibinfo{pages}{eadn5290}.
\newblock


\bibitem[Early and DeCosta-Smith(2011)]%
        {early2011making}
\bibfield{author}{\bibinfo{person}{Jessica~Singer Early} {and} \bibinfo{person}{Meredith DeCosta-Smith}.} \bibinfo{year}{2011}\natexlab{}.
\newblock \showarticletitle{Making a case for college: A genre-based college admission essay intervention for underserved high school students}.
\newblock \bibinfo{journal}{\emph{Journal of Writing Research}} \bibinfo{volume}{2}, \bibinfo{number}{3} (\bibinfo{year}{2011}), \bibinfo{pages}{299--329}.
\newblock


\bibitem[Early et~al\mbox{.}(2010)]%
        {early2010write}
\bibfield{author}{\bibinfo{person}{Jessica~Singer Early}, \bibinfo{person}{Meredith DeCosta-Smith}, {and} \bibinfo{person}{Arturo Valdespino}.} \bibinfo{year}{2010}\natexlab{}.
\newblock \showarticletitle{Write Your Ticket to College: A Genre-Based College Admission Essay Workshop for Ethnically Diverse, Underserved Students}.
\newblock \bibinfo{journal}{\emph{Journal of Adolescent \& Adult Literacy}} \bibinfo{volume}{54}, \bibinfo{number}{3} (\bibinfo{year}{2010}), \bibinfo{pages}{209--219}.
\newblock


\bibitem[Engzell et~al\mbox{.}(2021)]%
        {engzell2021learning}
\bibfield{author}{\bibinfo{person}{Per Engzell}, \bibinfo{person}{Arun Frey}, {and} \bibinfo{person}{Mark~D Verhagen}.} \bibinfo{year}{2021}\natexlab{}.
\newblock \showarticletitle{Learning loss due to school closures during the COVID-19 pandemic}.
\newblock \bibinfo{journal}{\emph{Proceedings of the national academy of sciences}} \bibinfo{volume}{118}, \bibinfo{number}{17} (\bibinfo{year}{2021}), \bibinfo{pages}{e2022376118}.
\newblock


\bibitem[Flesch(1948)]%
        {flesch1948new}
\bibfield{author}{\bibinfo{person}{Rudolf Flesch}.} \bibinfo{year}{1948}\natexlab{}.
\newblock \showarticletitle{A new readability yardstick}.
\newblock \bibinfo{journal}{\emph{Journal of Applied Psychology}} \bibinfo{volume}{32}, \bibinfo{number}{3} (\bibinfo{year}{1948}), \bibinfo{pages}{221--233}.
\newblock


\bibitem[Freeman(2025)]%
        {freeman2025student}
\bibfield{author}{\bibinfo{person}{Josh Freeman}.} \bibinfo{year}{2025}\natexlab{}.
\newblock \showarticletitle{Student generative ai survey 2025}.
\newblock \bibinfo{journal}{\emph{Higher Education Policy Institute: London, UK}} (\bibinfo{year}{2025}).
\newblock


\bibitem[Hargittai(2002)]%
        {hargittai2002second}
\bibfield{author}{\bibinfo{person}{Eszter Hargittai}.} \bibinfo{year}{2002}\natexlab{}.
\newblock \showarticletitle{Second-level digital divide: Differences in people's online skills}.
\newblock \bibinfo{journal}{\emph{First Monday}} \bibinfo{volume}{7}, \bibinfo{number}{4} (\bibinfo{year}{2002}).
\newblock


\bibitem[Hofmann et~al\mbox{.}(2024)]%
        {hofmann2024ai}
\bibfield{author}{\bibinfo{person}{Valentin Hofmann}, \bibinfo{person}{Pratyusha~Ria Kalluri}, \bibinfo{person}{Dan Jurafsky}, {and} \bibinfo{person}{Sharese King}.} \bibinfo{year}{2024}\natexlab{}.
\newblock \showarticletitle{AI generates covertly racist decisions about people based on their dialect}.
\newblock \bibinfo{journal}{\emph{Nature}} \bibinfo{volume}{633}, \bibinfo{number}{8028} (\bibinfo{year}{2024}), \bibinfo{pages}{147--154}.
\newblock


\bibitem[Huang(2024)]%
        {huang2024translating}
\bibfield{author}{\bibinfo{person}{Tiffany~J Huang}.} \bibinfo{year}{2024}\natexlab{}.
\newblock \showarticletitle{Translating authentic selves into authentic applications: Private college consulting and selective college admissions}.
\newblock \bibinfo{journal}{\emph{Sociology of Education}} \bibinfo{volume}{97}, \bibinfo{number}{2} (\bibinfo{year}{2024}), \bibinfo{pages}{174--192}.
\newblock


\bibitem[Hwang et~al\mbox{.}(2025)]%
        {hwang202580}
\bibfield{author}{\bibinfo{person}{Angel Hsing-Chi Hwang}, \bibinfo{person}{Q~Vera Liao}, \bibinfo{person}{Su~Lin Blodgett}, \bibinfo{person}{Alexandra Olteanu}, {and} \bibinfo{person}{Adam Trischler}.} \bibinfo{year}{2025}\natexlab{}.
\newblock \showarticletitle{'It was 80\% me, 20\% AI': Seeking Authenticity in Co-Writing with Large Language Models}.
\newblock \bibinfo{journal}{\emph{Proceedings of the ACM on Human-Computer Interaction}} \bibinfo{volume}{9}, \bibinfo{number}{2} (\bibinfo{year}{2025}), \bibinfo{pages}{1--41}.
\newblock


\bibitem[Imai et~al\mbox{.}(2010)]%
        {imai2010general}
\bibfield{author}{\bibinfo{person}{Kosuke Imai}, \bibinfo{person}{Luke Keele}, {and} \bibinfo{person}{Dustin Tingley}.} \bibinfo{year}{2010}\natexlab{}.
\newblock \showarticletitle{A general approach to causal mediation analysis}.
\newblock \bibinfo{journal}{\emph{Psychological Methods}} \bibinfo{volume}{15}, \bibinfo{number}{4} (\bibinfo{year}{2010}), \bibinfo{pages}{309--334}.
\newblock


\bibitem[J{\ae}ger and Karlson(2018)]%
        {jaeger2018cultural}
\bibfield{author}{\bibinfo{person}{Mads~Meier J{\ae}ger} {and} \bibinfo{person}{Kristian Karlson}.} \bibinfo{year}{2018}\natexlab{}.
\newblock \showarticletitle{Cultural capital and educational inequality: A counterfactual analysis}.
\newblock \bibinfo{journal}{\emph{Sociological Science}}  \bibinfo{volume}{5} (\bibinfo{year}{2018}), \bibinfo{pages}{775--795}.
\newblock


\bibitem[Jones(2013)]%
        {jones2013ensure}
\bibfield{author}{\bibinfo{person}{Steven Jones}.} \bibinfo{year}{2013}\natexlab{}.
\newblock \showarticletitle{“Ensure that you stand out from the crowd”: A corpus-based analysis of personal statements according to applicants’ school type}.
\newblock \bibinfo{journal}{\emph{Comparative Education Review}} \bibinfo{volume}{57}, \bibinfo{number}{3} (\bibinfo{year}{2013}), \bibinfo{pages}{397--423}.
\newblock


\bibitem[Kadoma et~al\mbox{.}(2024)]%
        {kadoma2024generative}
\bibfield{author}{\bibinfo{person}{Kowe Kadoma}, \bibinfo{person}{Dana{\"e} Metaxa}, {and} \bibinfo{person}{Mor Naaman}.} \bibinfo{year}{2024}\natexlab{}.
\newblock \showarticletitle{Generative AI and Perceptual Harms: Who's Suspected of using LLMs?}
\newblock \bibinfo{journal}{\emph{arXiv preprint arXiv:2410.00906}} (\bibinfo{year}{2024}).
\newblock


\bibitem[Karlson et~al\mbox{.}(2012)]%
        {karlson2012comparing}
\bibfield{author}{\bibinfo{person}{Kristian~Bernt Karlson}, \bibinfo{person}{Anders Holm}, {and} \bibinfo{person}{Richard Breen}.} \bibinfo{year}{2012}\natexlab{}.
\newblock \showarticletitle{Comparing coefficients of nested nonlinear probability models}.
\newblock \bibinfo{journal}{\emph{The Stata Journal}} \bibinfo{volume}{12}, \bibinfo{number}{3} (\bibinfo{year}{2012}), \bibinfo{pages}{420--438}.
\newblock


\bibitem[Khampusaen(2025)]%
        {khampusaen2025impact}
\bibfield{author}{\bibinfo{person}{Dararat Khampusaen}.} \bibinfo{year}{2025}\natexlab{}.
\newblock \showarticletitle{The Impact of ChatGPT on Academic Writing Skills and Knowledge: An Investigation of Its Use in Argumentative Essays.}
\newblock \bibinfo{journal}{\emph{LEARN Journal: Language Education and Acquisition Research Network}} \bibinfo{volume}{18}, \bibinfo{number}{1} (\bibinfo{year}{2025}), \bibinfo{pages}{963--988}.
\newblock


\bibitem[Kim et~al\mbox{.}(2023)]%
        {kim2023towards}
\bibfield{author}{\bibinfo{person}{Yewon Kim}, \bibinfo{person}{Mina Lee}, \bibinfo{person}{Donghwi Kim}, {and} \bibinfo{person}{Sung-Ju Lee}.} \bibinfo{year}{2023}\natexlab{}.
\newblock \showarticletitle{Towards explainable ai writing assistants for non-native english speakers}.
\newblock \bibinfo{journal}{\emph{arXiv preprint arXiv:2304.02625}} (\bibinfo{year}{2023}).
\newblock


\bibitem[Kleinberg and Raghavan(2021)]%
        {kleinberg2021algorithmic}
\bibfield{author}{\bibinfo{person}{Jon Kleinberg} {and} \bibinfo{person}{Manish Raghavan}.} \bibinfo{year}{2021}\natexlab{}.
\newblock \showarticletitle{Algorithmic monoculture and social welfare}.
\newblock \bibinfo{journal}{\emph{Proceedings of the National Academy of Sciences}} \bibinfo{volume}{118}, \bibinfo{number}{22} (\bibinfo{year}{2021}), \bibinfo{pages}{e2018340118}.
\newblock


\bibitem[Korchak et~al\mbox{.}(2025)]%
        {korchak2025enhancing}
\bibfield{author}{\bibinfo{person}{Anna Korchak}, \bibinfo{person}{Mik Fanguy}, \bibinfo{person}{Kseniia Adamovich}, \bibinfo{person}{Han Zhang}, \bibinfo{person}{Mattew Baldwin}, {and} \bibinfo{person}{Jamie Costley}.} \bibinfo{year}{2025}\natexlab{}.
\newblock \showarticletitle{Enhancing Collaborative Academic Writing with Generative Artificial Intelligence: Use and Effects}. In \bibinfo{booktitle}{\emph{International Conference on Artificial Intelligence in Education}}. Springer, \bibinfo{pages}{297--304}.
\newblock


\bibitem[Kosmyna et~al\mbox{.}(2025)]%
        {kosmyna2025your}
\bibfield{author}{\bibinfo{person}{Nataliya Kosmyna}, \bibinfo{person}{Eugene Hauptmann}, \bibinfo{person}{Ye~Tong Yuan}, \bibinfo{person}{Jessica Situ}, \bibinfo{person}{Xian-Hao Liao}, \bibinfo{person}{Ashly~Vivian Beresnitzky}, \bibinfo{person}{Iris Braunstein}, {and} \bibinfo{person}{Pattie Maes}.} \bibinfo{year}{2025}\natexlab{}.
\newblock \showarticletitle{Your brain on ChatGPT: Accumulation of cognitive debt when using an AI assistant for essay writing task}.
\newblock \bibinfo{journal}{\emph{arXiv preprint arXiv:2506.08872}}  \bibinfo{volume}{4} (\bibinfo{year}{2025}).
\newblock


\bibitem[Kusumegi et~al\mbox{.}(2025)]%
        {kusumegi2025scientific}
\bibfield{author}{\bibinfo{person}{Keigo Kusumegi}, \bibinfo{person}{Xinyu Yang}, \bibinfo{person}{Paul Ginsparg}, \bibinfo{person}{Mathijs de Vaan}, \bibinfo{person}{Toby Stuart}, {and} \bibinfo{person}{Yian Yin}.} \bibinfo{year}{2025}\natexlab{}.
\newblock \showarticletitle{Scientific production in the era of large language models}.
\newblock \bibinfo{journal}{\emph{Science}} \bibinfo{volume}{390}, \bibinfo{number}{6779} (\bibinfo{year}{2025}), \bibinfo{pages}{1240--1243}.
\newblock


\bibitem[Lee et~al\mbox{.}(2025)]%
        {lee2025poor}
\bibfield{author}{\bibinfo{person}{Jinsook Lee}, \bibinfo{person}{AJ Alvero}, \bibinfo{person}{Thorsten Joachims}, {and} \bibinfo{person}{Rene Kizilcec}.} \bibinfo{year}{2025}\natexlab{}.
\newblock \showarticletitle{Poor Alignment and Steerability of Large Language Models: Evidence from College Admission Essays}.
\newblock \bibinfo{journal}{\emph{arXiv preprint arXiv:2503.20062}} (\bibinfo{year}{2025}).
\newblock


\bibitem[Liang et~al\mbox{.}(2024)]%
        {liangmonitoring}
\bibfield{author}{\bibinfo{person}{Weixin Liang}, \bibinfo{person}{Zachary Izzo}, \bibinfo{person}{Yaohui Zhang}, \bibinfo{person}{Haley Lepp}, \bibinfo{person}{Hancheng Cao}, \bibinfo{person}{Xuandong Zhao}, \bibinfo{person}{Lingjiao Chen}, \bibinfo{person}{Haotian Ye}, \bibinfo{person}{Sheng Liu}, \bibinfo{person}{Zhi Huang}, {et~al\mbox{.}}} \bibinfo{year}{2024}\natexlab{}.
\newblock \showarticletitle{Monitoring AI-Modified Content at Scale: A Case Study on the Impact of ChatGPT on AI Conference Peer Reviews}. In \bibinfo{booktitle}{\emph{Forty-first International Conference on Machine Learning}}.
\newblock


\bibitem[Maas(1972)]%
        {maas1972uber}
\bibfield{author}{\bibinfo{person}{Heinz-Dieter Maas}.} \bibinfo{year}{1972}\natexlab{}.
\newblock \showarticletitle{{\"U}ber den Zusammenhang zwischen Wortschatzumfang und L{\"a}nge eines Textes}.
\newblock \bibinfo{journal}{\emph{Zeitschrift f{\"u}r Literaturwissenschaft und Linguistik}} \bibinfo{volume}{2}, \bibinfo{number}{8} (\bibinfo{year}{1972}), \bibinfo{pages}{73--79}.
\newblock


\bibitem[McCarthy and Jarvis(2007)]%
        {mccarthy2007vocd}
\bibfield{author}{\bibinfo{person}{Philip~M. McCarthy} {and} \bibinfo{person}{Scott Jarvis}.} \bibinfo{year}{2007}\natexlab{}.
\newblock \showarticletitle{vocd: A theoretical and empirical evaluation}.
\newblock \bibinfo{journal}{\emph{Language Testing}} \bibinfo{volume}{24}, \bibinfo{number}{4} (\bibinfo{year}{2007}), \bibinfo{pages}{459--488}.
\newblock


\bibitem[McCarthy and Jarvis(2010)]%
        {mccarthy2010mtld}
\bibfield{author}{\bibinfo{person}{Philip~M. McCarthy} {and} \bibinfo{person}{Scott Jarvis}.} \bibinfo{year}{2010}\natexlab{}.
\newblock \showarticletitle{{MTLD}, vocd-{D}, and {HD-D}: A validation study of sophisticated approaches to lexical diversity assessment}.
\newblock \bibinfo{journal}{\emph{Behavior Research Methods}} \bibinfo{volume}{42}, \bibinfo{number}{2} (\bibinfo{year}{2010}), \bibinfo{pages}{381--392}.
\newblock


\bibitem[Moon et~al\mbox{.}(2024)]%
        {moon2024homogenizing}
\bibfield{author}{\bibinfo{person}{Kibum Moon}, \bibinfo{person}{Adam Green}, {and} \bibinfo{person}{Kostadin Kushlev}.} \bibinfo{year}{2024}\natexlab{}.
\newblock \bibinfo{title}{Homogenizing Effect of Large Language Model (LLM) on Creative Diversity: An Empirical Comparison}.
\newblock


\bibitem[Myung et~al\mbox{.}(2025)]%
        {myung2025scaffolding}
\bibfield{author}{\bibinfo{person}{Junho Myung}, \bibinfo{person}{Hyunseung Lim}, \bibinfo{person}{Hana Oh}, \bibinfo{person}{Hyoungwook Jin}, \bibinfo{person}{Nayeon Kang}, \bibinfo{person}{So-Yeon Ahn}, \bibinfo{person}{Hwajung Hong}, \bibinfo{person}{Alice Oh}, {and} \bibinfo{person}{Juho Kim}.} \bibinfo{year}{2025}\natexlab{}.
\newblock \showarticletitle{When Scaffolding Breaks: Investigating Student Interaction with LLM-Based Writing Support in Real-Time K-12 EFL Classrooms}.
\newblock \bibinfo{journal}{\emph{arXiv preprint arXiv:2512.05506}} (\bibinfo{year}{2025}).
\newblock


\bibitem[Nash(1990)]%
        {nash1990bourdieu}
\bibfield{author}{\bibinfo{person}{Roy Nash}.} \bibinfo{year}{1990}\natexlab{}.
\newblock \showarticletitle{Bourdieu on education and social and cultural reproduction}.
\newblock \bibinfo{journal}{\emph{British journal of sociology of education}} \bibinfo{volume}{11}, \bibinfo{number}{4} (\bibinfo{year}{1990}), \bibinfo{pages}{431--447}.
\newblock


\bibitem[Nghiem et~al\mbox{.}(2025)]%
        {nghiem2025rich}
\bibfield{author}{\bibinfo{person}{Huy Nghiem}, \bibinfo{person}{Phuong-Anh Nguyen-Le}, \bibinfo{person}{John Prindle}, \bibinfo{person}{Rachel Rudinger}, {and} \bibinfo{person}{Hal Daum{\'e}~III}.} \bibinfo{year}{2025}\natexlab{}.
\newblock \showarticletitle{‘Rich Dad, Poor Lad’: How do Large Language Models Contextualize Socioeconomic Factors in College Admission?}. In \bibinfo{booktitle}{\emph{Proceedings of the 2025 Conference on Empirical Methods in Natural Language Processing}}. \bibinfo{pages}{21033--21067}.
\newblock


\bibitem[Noy and Zhang(2023)]%
        {noy2023experimental}
\bibfield{author}{\bibinfo{person}{Shakked Noy} {and} \bibinfo{person}{Whitney Zhang}.} \bibinfo{year}{2023}\natexlab{}.
\newblock \showarticletitle{Experimental evidence on the productivity effects of generative artificial intelligence}.
\newblock \bibinfo{journal}{\emph{Science}} \bibinfo{volume}{381}, \bibinfo{number}{6654} (\bibinfo{year}{2023}), \bibinfo{pages}{187--192}.
\newblock


\bibitem[OpenAI(2024)]%
        {openai2024gpt4o}
\bibfield{author}{\bibinfo{person}{OpenAI}.} \bibinfo{year}{2024}\natexlab{}.
\newblock \bibinfo{title}{GPT-4 Open System Card}.
\newblock
\urldef\tempurl%
\url{https://cdn.openai.com/gpt-4o-system-card.pdf}
\showURL{%
\tempurl}
\newblock
\shownote{Accessed: December 2, 2024}.


\bibitem[Polakova and Ivenz(2024)]%
        {polakova2024impact}
\bibfield{author}{\bibinfo{person}{Petra Polakova} {and} \bibinfo{person}{Petra Ivenz}.} \bibinfo{year}{2024}\natexlab{}.
\newblock \showarticletitle{The impact of ChatGPT feedback on the development of EFL students’ writing skills}.
\newblock \bibinfo{journal}{\emph{Cogent Education}} \bibinfo{volume}{11}, \bibinfo{number}{1} (\bibinfo{year}{2024}), \bibinfo{pages}{2410101}.
\newblock


\bibitem[Sabnis et~al\mbox{.}(2022)]%
        {sabnis2022large}
\bibfield{author}{\bibinfo{person}{Sunil Sabnis}, \bibinfo{person}{Renzhe Yu}, {and} \bibinfo{person}{Ren{\'e}~F Kizilcec}.} \bibinfo{year}{2022}\natexlab{}.
\newblock \showarticletitle{Large-scale student data reveal sociodemographic gaps in procrastination behavior}. In \bibinfo{booktitle}{\emph{Proceedings of the ninth ACM conference on learning@ scale}}. \bibinfo{pages}{133--141}.
\newblock


\bibitem[Stevens(2009)]%
        {stevens2009creating}
\bibfield{author}{\bibinfo{person}{Mitchell~L Stevens}.} \bibinfo{year}{2009}\natexlab{}.
\newblock \bibinfo{booktitle}{\emph{Creating a class}}.
\newblock \bibinfo{publisher}{Harvard University Press}.
\newblock


\bibitem[Stofiana et~al\mbox{.}(2025)]%
        {stofiana2025writing}
\bibfield{author}{\bibinfo{person}{Tofan Stofiana}, \bibinfo{person}{Dadang Sunendar}, \bibinfo{person}{Yeti Mulyati}, {and} \bibinfo{person}{Andoyo Sastromiharjo}.} \bibinfo{year}{2025}\natexlab{}.
\newblock \showarticletitle{Writing with AI, thinking with Toulmin: metacognitive gaps and the rhetorical limits of argumentation}.
\newblock \bibinfo{journal}{\emph{Ampersand}} (\bibinfo{year}{2025}), \bibinfo{pages}{100242}.
\newblock


\bibitem[Templin(1957)]%
        {templin1957certain}
\bibfield{author}{\bibinfo{person}{Mildred~C. Templin}.} \bibinfo{year}{1957}\natexlab{}.
\newblock \bibinfo{booktitle}{\emph{Certain Language Skills in Children: Their Development and Interrelationships}}.
\newblock \bibinfo{publisher}{University of Minnesota Press}, \bibinfo{address}{Minneapolis, MN}.
\newblock


\bibitem[Tingley et~al\mbox{.}(2014)]%
        {tingley2014mediation}
\bibfield{author}{\bibinfo{person}{Dustin Tingley}, \bibinfo{person}{Teppei Yamamoto}, \bibinfo{person}{Kentaro Hirose}, \bibinfo{person}{Luke Keele}, {and} \bibinfo{person}{Kosuke Imai}.} \bibinfo{year}{2014}\natexlab{}.
\newblock \showarticletitle{mediation: {R} Package for Causal Mediation Analysis}.
\newblock \bibinfo{journal}{\emph{Journal of Statistical Software}} \bibinfo{volume}{59}, \bibinfo{number}{5} (\bibinfo{year}{2014}), \bibinfo{pages}{1--38}.
\newblock


\bibitem[VanderWeele(2015)]%
        {vanderweele2015explanation}
\bibfield{author}{\bibinfo{person}{Tyler~J VanderWeele}.} \bibinfo{year}{2015}\natexlab{}.
\newblock \bibinfo{booktitle}{\emph{Explanation in Causal Inference: Methods for Mediation and Interaction}}.
\newblock \bibinfo{publisher}{Oxford University Press}.
\newblock


\bibitem[Viberg et~al\mbox{.}(2024)]%
        {viberg2024advancing}
\bibfield{author}{\bibinfo{person}{Olga Viberg}, \bibinfo{person}{Ren{\'e}~F Kizilcec}, \bibinfo{person}{Alyssa~Friend Wise}, \bibinfo{person}{Ioana Jivet}, {and} \bibinfo{person}{Nia Nixon}.} \bibinfo{year}{2024}\natexlab{}.
\newblock \bibinfo{title}{Advancing equity and inclusion in educational practices with AI-powered educational decision support systems (AI-EDSS)}.
\newblock \bibinfo{numpages}{1974--1981}~pages.
\newblock


\bibitem[Warren(2013)]%
        {warren2013rhetoric}
\bibfield{author}{\bibinfo{person}{James Warren}.} \bibinfo{year}{2013}\natexlab{}.
\newblock \showarticletitle{The rhetoric of college application essays: Removing obstacles for low income and minority students}.
\newblock \bibinfo{journal}{\emph{American Secondary Education}} (\bibinfo{year}{2013}), \bibinfo{pages}{43--56}.
\newblock


\bibitem[Warschauer(2004)]%
        {warschauer2004technology}
\bibfield{author}{\bibinfo{person}{Mark Warschauer}.} \bibinfo{year}{2004}\natexlab{}.
\newblock \bibinfo{booktitle}{\emph{Technology and social inclusion: Rethinking the digital divide}}.
\newblock \bibinfo{publisher}{MIT press}.
\newblock


\bibitem[Yan et~al\mbox{.}(2024)]%
        {yan2024practical}
\bibfield{author}{\bibinfo{person}{Lixiang Yan}, \bibinfo{person}{Lele Sha}, \bibinfo{person}{Linxuan Zhao}, \bibinfo{person}{Yuheng Li}, \bibinfo{person}{Roberto Martinez-Maldonado}, \bibinfo{person}{Guanliang Chen}, \bibinfo{person}{Xinyu Li}, \bibinfo{person}{Yueqiao Jin}, {and} \bibinfo{person}{Dragan Ga{\v{s}}evi{\'c}}.} \bibinfo{year}{2024}\natexlab{}.
\newblock \showarticletitle{Practical and ethical challenges of large language models in education: A systematic scoping review}.
\newblock \bibinfo{journal}{\emph{British Journal of Educational Technology}} \bibinfo{volume}{55}, \bibinfo{number}{1} (\bibinfo{year}{2024}), \bibinfo{pages}{90--112}.
\newblock


\bibitem[Yu et~al\mbox{.}(2024)]%
        {yu2024whose}
\bibfield{author}{\bibinfo{person}{Renzhe Yu}, \bibinfo{person}{Zhen Xu}, \bibinfo{person}{Sky CH-Wang}, {and} \bibinfo{person}{Richard Arum}.} \bibinfo{year}{2024}\natexlab{}.
\newblock \showarticletitle{Whose chatgpt? unveiling real-world educational inequalities introduced by large language models}.
\newblock \bibinfo{journal}{\emph{arXiv preprint arXiv:2410.22282}} (\bibinfo{year}{2024}).
\newblock


\bibitem[Yule(1944)]%
        {yule1944statistical}
\bibfield{author}{\bibinfo{person}{George~Udny Yule}.} \bibinfo{year}{1944}\natexlab{}.
\newblock \bibinfo{booktitle}{\emph{The Statistical Study of Literary Vocabulary}}.
\newblock \bibinfo{publisher}{Cambridge University Press}, \bibinfo{address}{Cambridge}.
\newblock


\bibitem[Zdravkova and Ilijoski(2025)]%
        {zdravkova2025impact}
\bibfield{author}{\bibinfo{person}{Katerina Zdravkova} {and} \bibinfo{person}{Bojan Ilijoski}.} \bibinfo{year}{2025}\natexlab{}.
\newblock \showarticletitle{The impact of large language models on computer science student writing}.
\newblock \bibinfo{journal}{\emph{International Journal of Educational Technology in Higher Education}} \bibinfo{volume}{22}, \bibinfo{number}{1} (\bibinfo{year}{2025}), \bibinfo{pages}{32}.
\newblock


\bibitem[Zhang et~al\mbox{.}({[n.\,d.]})]%
        {zhang2025generative}
\bibfield{author}{\bibinfo{person}{Simone Zhang}, \bibinfo{person}{Janet Xu}, {and} \bibinfo{person}{AJ Alvero}.} \bibinfo{year}{[n.\,d.]}\natexlab{}.
\newblock \showarticletitle{Generative AI meets open-ended survey responses: Research participant use of ai and homogenization}.
\newblock \bibinfo{journal}{\emph{Sociological Methods \& Research}} (\bibinfo{year}{[n.\,d.]}), \bibinfo{pages}{00491241251327130}.
\newblock


\end{thebibliography}

\appendix
\newpage
\onecolumn
\section{Details of the Case Institution}\label{appdx:case_inst}

Our case institution is a highly selective, engineering-focused university in the United States that requires first-year applicants to apply through the Common App. Through the Common App, the institution collects a wide range of applicant information, including standardized test scores (e.g., SAT), high school grades and coursework, and demographic characteristics. It also collects multiple forms of unstructured text, including applicant essays and recommendation letters.

For this study, we analyzed first-year applications to the engineering school from the 2019–2020 through 2023–2024 admissions cycles. We focused on essays written in response to eight Common App prompts (Table~\ref{common_app_essay}), yielding 81,663 applications. Essays were limited to 650 words. We excluded essays shorter than 250 words, as such essays were not considered in the institution’s review process.

\begin{table*}[!htpb]
\centering
\caption{Common App Essay Questions and Their Frequency in the Human Essay Sample.}
\label{common_app_essay}
\begin{tabular}{p{0.9\textwidth}c}
\toprule
\textbf{Essay Question} & \textbf{Count} \\ 
\midrule
Share an essay on any topic of your choice. It can be one you've already written, one that responds to a different prompt, or one of your own design. & 19,768\\
\midrule
Discuss an accomplishment, event, or realization that sparked a period of personal growth and a new understanding of yourself or others. & 19,557 \\
\midrule
Some students have a background, identity, interest, or talent that is so meaningful they believe their application would be incomplete without it. If this sounds like you, then please share your story. & 18,781 \\
\midrule
The lessons we take from obstacles we encounter can be fundamental to later success. Recount a time when you faced a challenge, setback, or failure. How did it affect you, and what did you learn from the experience? & 14,281\\
\midrule
Describe a topic, idea, or concept you find so engaging that it makes you lose all track of time. Why does it captivate you? What or who do you turn to when you want to learn more? & 4,449\\
\midrule
Reflect on a time when you questioned or challenged a belief or idea. What prompted your thinking? What was the outcome? & 2,149 \\
\midrule
Describe a problem you've solved or a problem you'd like to solve. It can be an intellectual challenge, a research query, an ethical dilemma-anything that is of personal importance, no matter the scale. Explain its significance to you and what steps you took or could be taken to identify a solution. & 1,454 \\
\midrule
Reflect on something that someone has done for you that has made you happy or thankful in a surprising way. How has this gratitude affected or motivated you? &1,222 \\
\bottomrule
\end{tabular}
\end{table*}

\begin{table}[!htpb]
\centering
\caption{An example prompt to generate a synthetic college admissions essay. The \colorbox{green!15}{highlighted} portion is modified based on the actual distribution of essay prompt choices.}
\label{tab:llm_prompt}
\small
\ttfamily
\begin{tabular}{p{\columnwidth}}
\toprule
\textbf{User Prompt}\\
\midrule
I am a high school student applying to the [case institution]'s College of Engineering. Here is a little bit more information about myself. \\[6pt]
\\
I have to write an essay as part of my application. The essay must be longer than 250 words but no more than 650 words. Below are the instructions for writing the essay and the specific prompt I need to respond to. Write an essay based on the given instructions and prompt. \\[6pt]
\\
Instructions: "The essay demonstrates your ability to write clearly and concisely on a selected topic and helps you distinguish yourself in your own voice. What do you want the readers of your application to know about you apart from courses, grades, and test scores? Choose the option that best helps you answer that question and write an essay of no more than 650 words, using the prompt to inspire and structure your response.\\
\\
Remember: 650 words is your limit, not your goal. Use the full range if you need it, but don't feel obligated to do so. (The application won't accept a response shorter than 250 words.)" \\[6pt]
\\
\colorbox{green!15}{\parbox{\dimexpr\columnwidth-2\fboxsep}{Prompt: "\textit{Share an essay on any topic of your choice. It can be one you've already written, one that responds to a different prompt, or one of your own design.}"}} \\
\bottomrule
\end{tabular}
\end{table}

\section{Distribution of LLM Estimation Result and Validation}\label{appdx:llm_distribution_validation}

\begin{figure}[!htbp]
    \centering
    \includegraphics[width=0.9\linewidth]{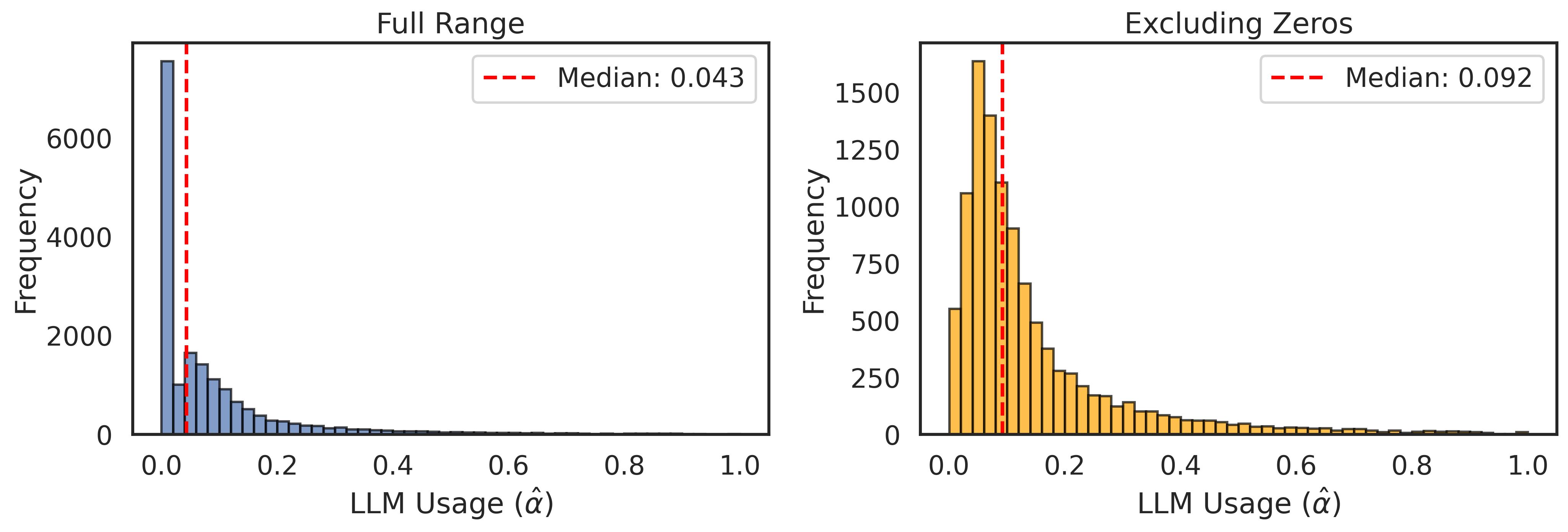}
    \caption{Distribution of $\hat{\alpha}$} 
    \label{fig:alpha_distribution}
\end{figure}

Figure \ref{fig:alpha_distribution} presents the distribution of estimated LLM usage ($\hat{\alpha}$) among post-GPT applicants. A large mass at exactly zero, representing applicants with no detectable LLM involvement, and a right-skewed continuous distribution among users. The overall median is $\hat{\alpha} = 0.043$. The right panel shows the distribution conditional on any detected usage ($\hat{\alpha} > 0$), which follows a right-skewed unimodal distribution with a median of $\hat{\alpha} = 0.092$. Most users cluster in the low-to-moderate range ($0.02 < \hat{\alpha} < 0.20$), with a long tail extending toward higher values. The clean separation between zero and nonzero values supports our stratification approach, which classifies applicants with $\hat{\alpha} = 0$ as non-users and divides the remaining applicants into terciles of estimated usage.

\begin{figure}[!htbp]
    \centering
    \includegraphics[width=0.5\linewidth]{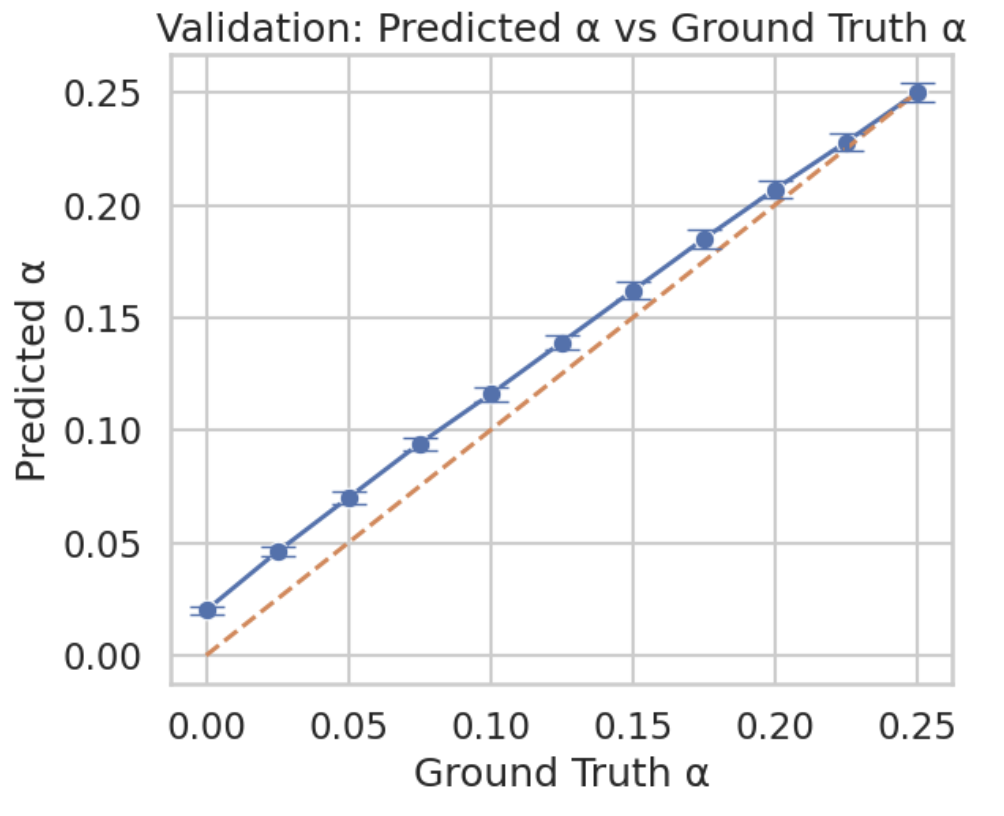}
    \caption{Validation of LLM-usage estimator} 
    \label{fig:MLE_validation}
\end{figure}

Figure \ref{fig:MLE_validation} shows validation test results of LLM estimation. Predicted mixing proportion $\alpha$ ($\hat{\alpha}$) versus ground-truth $\alpha$ on a held-out validation set with known mixture fractions. Points show the mean predicted $\alpha$ within each ground truth bin, and error bars indicate 95\% confidence intervals. The dashed line denotes perfect calibration (Predicted $\alpha$ = $\alpha$). The estimator tracks the identity line closely, with a small positive baseline at $\hat{\alpha}$ = 0 consistent with the estimation error.

\section{Formal Assumption Checks for Difference in Difference Analysis}\label{appdx:did_formal_check}

\subsection{Check 1. Parallel Trends.} To evaluate the validity of the parallel trends assumption underlying our difference-in-differences framework, we estimate an event study specification interacting the SES indicator with year fixed effects, using the 2023 admission cycle as the reference period. If the parallel trends assumption holds, the interaction coefficients for pre-treatment cycles should be statistically indistinguishable from zero, indicating that both SES group applicants followed similar trajectories in the outcome prior to the introduction of ChatGPT.

Table \ref{tab:event_study} shows that the 2020 and 2021 interaction coefficients are small and not statistically significant ($\hat{\beta}_{2020} = -0.050$, $p = 0.782$; $\hat{\beta}_{2021} = 0.241$, $p = 0.145$), consistent with parallel pre-trends in those cycles. However, the 2022 interaction term is substantively larger and statistically significant ($\hat{\beta}_{2022} = 0.476$, $p = 0.004$), and a joint Wald test rejects the null that all pre-treatment interactions are simultaneously zero ($p = 0.006$). The post-treatment coefficient for 2024 is near zero and nonsignificant ($\hat{\beta}_{2024} = 0.008$, $p = 0.948$).

\begin{figure}[!htbp]
    \centering
    \includegraphics[width=0.7\linewidth]{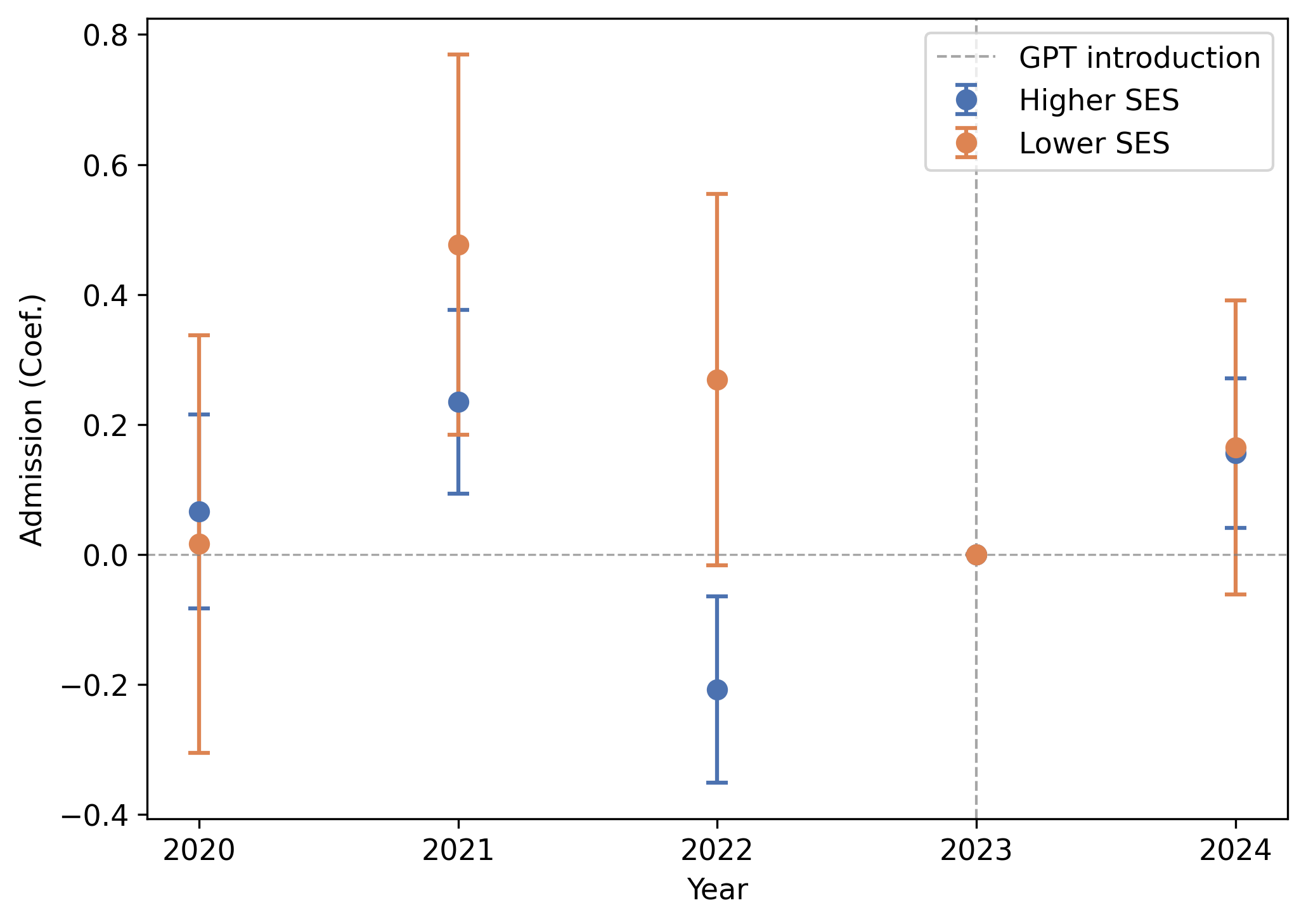}
    \caption{Coefficient estimates and 95\% confidence intervals for logistic regressions estimating the treatment effect of the LLM introduction on college admission for SES groups.} 
    \label{fig:coef_event_study}
\end{figure}

The divergence in 2022 likely reflects disruptions introduced by the COVID-19 pandemic and the widespread adoption of test-optional admissions policies during that period (See \ref{fig:coef_event_study}. These policy shifts disproportionately altered the composition and behavior of applicant pools in ways that varied across socioeconomic groups, producing differential pre-treatment trajectories that are unrelated to LLM availability. Because the parallel trends assumption is sensitive to this pre-treatment divergence, we interpret our difference-in-differences estimates as descriptive rather than causal. The results characterize how the association between SES status and admission outcomes shifted following LLMs' introduction, but we do not claim that these shifts are causally attributable to LLM adoption. 

\begin{table}[htbp]
\centering
\caption{Event Study: Fee Waiver $\times$ Year Interactions (Logit)}
\label{tab:event_study}
\begin{tabular}{lccccc}
\toprule
\textbf{Year} & \textbf{Coefficient} & \textbf{Std.\ Error} & \textbf{95\% CI} & \textbf{$p$-value }& \textbf{Pre-treatment} \\
\midrule
2020 & $-0.050$ & $(0.180)$ & $[-0.404,\; 0.304]$ & $0.782$ & \checkmark \\
2021 & $0.241$  & $(0.166)$ & $[-0.083,\; 0.566]$ & $0.145$ & \checkmark \\
2022 & $0.476^{**}$ & $(0.163)$ & $[0.157,\; 0.796]$ & $0.004$ & \checkmark \\
2023 (ref.) & $0.000$ & --- & --- & --- & \checkmark \\
2024 & $0.008$ & $(0.129)$ & $[-0.245,\; 0.262]$ & $0.948$ &  \\
\midrule
\multicolumn{6}{l}{Joint Wald test (pre-treatment interactions $= 0$): $\chi^2 = [\cdot],\; p = 0.006$} \\
\bottomrule
\end{tabular}

\vspace{0.5em}
\begin{minipage}{0.92\textwidth}
\footnotesize
\textit{Notes:} Table reports coefficients from a logit model interacting the fee waiver indicator with admission cycle fixed effects. The 2023 cycle serves as the reference period. Pre-treatment cycles (2020--2022) precede the public release of ChatGPT. The 2022 divergence likely reflects disruptions from COVID-19 pandemic test-optional policies. Significance levels: $^{*}\,p<0.05$, $^{**}\,p<0.01$, $^{***}\,p<0.001$.
\end{minipage}
\end{table}

\subsection{Check 2. Placebo Treatment Timing}
As a further diagnostic, we conduct a placebo treatment timing test in which we assign counterfactual treatment dates to pre-GPT period and estimate the Fee Waver $\times$ Post interaction under each. If the differential patterns we observe at the true treatment date reflect pre-existing diverging trends rather than a shift associated with ChatGPT's introduction, we would expect significant interaction effects at placebo treatment dates as well. Table~\ref{tab:placebo} reports the results. None of the three placebo specifications produces a statistically significant interaction: assigning fake treatment at 2021 ($\hat{\beta} = 0.230$, $p = 0.124$), 2022 ($\hat{\beta} = 0.113$, $p = 0.339$), and 2023 ($\hat{\beta} = -0.257$, $p = 0.062$) all fail to reject the null of no differential effect. The 2023 placebo approaches marginal significance but carries the opposite sign which suggests it does not reflect a confounding upward trend.

These results complement the event study findings. While the event study identified a significant pre-treatment divergence in the 2022 cycle, the placebo timing test finds no evidence that imposing a binary treatment split at any pre-period cutoff produces a significant differential effect. This pattern is consistent with the interpretation that the 2022 divergence reflects a discrete, period-specific disruption, plausibly attributable to COVID-19 pandemic test-optional policies, rather than a sustained differential trend between fee-waiver and non-fee-waiver applicants. Thus, the event study and placebo results motivate our decision to interpret the difference-in-differences estimates as descriptive of shifting associations rather than as evidence of a causal effect of LLM adoption on admission outcomes.

\begin{table}[htbp]
\centering
\caption{Placebo Treatment Timing Test (Logit)}
\label{tab:placebo}
\begin{tabular}{lccccl}
\toprule
\textbf{Fake Treatment Cycle} & \textbf{Coefficient} & \textbf{Std.\ Error} & \textbf{95\% CI }& $p$-value & \textbf{Significant} \\
\midrule
2021 & $0.230$ & $(0.149)$ & $[-0.063,\; 0.523]$ & $0.124$ & No \\
2022 & $0.113$ & $(0.118)$ & $[-0.119,\; 0.346]$ & $0.339$ & No \\
2023 & $-0.257$ & $(0.138)$ & $[-0.527,\; 0.013]$ & $0.062$ & No \\
\bottomrule
\end{tabular}

\vspace{0.5em}
\begin{minipage}{0.92\textwidth}
\footnotesize
\textit{Notes:} Table reports the coefficient on the fee waiver $\times$ fake post interaction from logit models estimated on the pre-treatment sample (cycles 2020--2023). Each row assigns a different counterfactual treatment date. The dependent variable is admission. All specifications include confounders. None of the placebo interactions reaches statistical significance at the 5\% level. Significance levels: $^{*}\,p<0.05$, $^{**}\,p<0.01$, $^{***}\,p<0.001$.
\end{minipage}
\end{table}

\subsection{Check 3. COVID sensitivity}
\paragraph{COVID Interaction.}
To test whether COVID-era disruptions directly confound the DiD estimate, we augment the model with a COVID-era indicator (equal to one for the 2020 and 2021 cycles) and its interaction with fee-waiver status (Table~\ref{tab:covid_interaction}). Adding the COVID-era fixed effect leaves the DiD coefficient largely unchanged ($\hat{\beta} = -0.181$, $p = 0.017$, compared to $\hat{\beta} = -0.170$, $p = 0.025$ in the base model). Including the COVID $\times$ fee waiver interaction produces a nonsignificant interaction term ($\hat{\beta}_{\text{COVID} \times \text{FW}} = -0.143$, $p = 0.229$), indicating that COVID did not differentially affect fee-waiver and non-fee-waiver applicants in a manner that accounts for the aggregate DiD estimate. However, the DiD coefficient in this specification ($\hat{\beta} = -0.248$, $p = 0.008$) should be interpreted cautiously, as subsequent tests reveal its sensitivity to sample composition.

\begin{table}[htbp]
\centering
\caption{COVID Interaction Test (Logit)}
\label{tab:covid_interaction}
\begin{tabular}{lcccc}
\toprule
\textbf{Specification} &\textbf{ DiD Coefficient} & \textbf{Std.\ Error} & \textbf{$p$-value} & \textbf{COVID $\times$ SES} \\
\midrule
Base DiD               & $-0.170$ & $(0.076)$ & $0.025$ & \\
+ COVID era FE         & $-0.181$ & $(0.076)$ & $0.017$ & \\
+ COVID $\times$ SES    & $-0.248$ & $(0.094)$ & $0.008$ & $-0.143\;(p = 0.229)$ \\
\bottomrule
\end{tabular}

\vspace{0.5em}
\begin{minipage}{0.92\textwidth}
\footnotesize
\textit{Notes:} Table reports the fee waiver $\times$ post coefficient from logit models with progressively richer COVID controls. The COVID era indicator equals one for the 2020 and 2021 admission cycles. The COVID $\times$ FW interaction is nonsignificant, indicating that pandemic-era disruptions did not differentially affect fee-waiver applicants in a manner that confounds the DiD estimate. All specifications include the full set of confounders.
\end{minipage}
\end{table}

\paragraph{Rolling Window.}
We estimate the DiD on each consecutive pair of admission cycles to identify which year-to-year transitions drive the aggregate result (Table~\ref{tab:rolling}). The only significant window is 2022 $\rightarrow$ 2023 ($\hat{\beta} = -0.494$, $p = 0.003$). All other transitions are nonsignificant: 2020 $\rightarrow$ 2021 ($\hat{\beta} = 0.212$, $p = 0.225$), 2021 $\rightarrow$ 2022 ($\hat{\beta} = 0.233$, $p = 0.141$), and critically, 2023 $\rightarrow$ 2024 ($\hat{\beta} = 0.008$, $p = 0.949$). The transition that most directly corresponds to the introduction of ChatGPT, from the 2023 to the 2024 admission cycle, shows no differential effect between fee-waiver and non-fee-waiver applicants. The aggregate DiD result is driven entirely by the large negative shift between 2022 and 2023.

\begin{table}[htbp]
\centering
\caption{Rolling Window Difference-in-Differences (Logit)}
\label{tab:rolling}
\begin{tabular}{lcccc}
\toprule
\textbf{Window} & \textbf{Coefficient} & \textbf{Std.\ Error} &\textbf{ $p$-value} & \textbf{$N$} \\
\midrule
2020 $\rightarrow$ 2021 & $0.212$       & $(0.175)$ & $0.225$ & \\
2021 $\rightarrow$ 2022 & $0.233$       & $(0.158)$ & $0.141$ & \\
2022 $\rightarrow$ 2023 & $-0.494^{**}$ & $(0.164)$ & $0.003$ & \\
2023 $\rightarrow$ 2024 & $0.008$       & $(0.130)$ & $0.949$ & \\
\bottomrule
\end{tabular}

\vspace{0.5em}
\begin{minipage}{0.92\textwidth}
\footnotesize
\textit{Notes:} Table reports the fee waiver $\times$ post coefficient from logit models estimated on each consecutive pair of admission cycles. The only significant transition is 2022 $\rightarrow$ 2023. The 2023 $\rightarrow$ 2024 window, corresponding to the introduction of ChatGPT, shows no differential effect. All specifications include the full set of confounders. Significance levels: $^{*}\,p<0.05$, $^{**}\,p<0.01$, $^{***}\,p<0.001$.
\end{minipage}
\end{table}

\paragraph{Donut-Hole Estimation.}
Finally, we systematically remove potentially contaminated cycles and re-estimate the DiD (Table~\ref{tab:donut}). The full sample estimate is significant ($\hat{\beta} = -0.170$, $p = 0.025$), and significance persists when dropping 2020 ($\hat{\beta} = -0.229$, $p = 0.005$) or both 2020 and 2021 ($\hat{\beta} = -0.252$, $p = 0.008$). However, when 2021 and 2022 are excluded, the estimate collapses ($\hat{\beta} = 0.042$, $p = 0.678$). Dropping all pre-2023 cycles yields an estimate indistinguishable from zero ($\hat{\beta} = 0.008$, $p = 0.949$), as does the ``clean ends'' specification comparing only 2020 and 2024 ($\hat{\beta} = 0.080$, $p = 0.577$). The significant aggregate estimates survive only when the 2022 cycle is included in the pre-period, confirming that the result is driven by the 2022$\rightarrow$2023 transition rather than by a post-treatment shift.

\begin{table}[htbp]
\centering
\caption{Donut-Hole Estimation: Sensitivity to Cycle Exclusion (Logit)}
\label{tab:donut}
\begin{tabular}{lcccl}
\toprule
\textbf{Specification} & \textbf{Coefficient} & \textbf{$p$-value} & \textbf{$N$} & \textbf{Cycles Included} \\
\midrule
Full sample              & $-0.170^{*}$  & $0.025$ & $30{,}488$ & 2020--2024 \\
Drop 2020                & $-0.229^{**}$ & $0.005$ & $27{,}840$ & 2021--2024 \\
Drop 2020--2021          & $-0.252^{**}$ & $0.008$ & $24{,}505$ & 2022--2024 \\
Drop 2021--2022          & $0.042$       & $0.678$ & $23{,}297$ & 2020, 2023--2024 \\
Drop 2020--2022          & $0.008$       & $0.949$ & $20{,}649$ & 2023--2024 \\
Clean ends: 2020 vs 2024 & $0.080$       & $0.577$ & $20{,}302$ & 2020, 2024 \\
\bottomrule
\end{tabular}

\vspace{0.5em}
\begin{minipage}{0.92\textwidth}
\footnotesize
\textit{Notes:} Table reports the fee waiver $\times$ post coefficient from logit models estimated on restricted samples. Significant estimates survive only when the 2022 cycle is included in the pre-period. Excluding 2021--2022, excluding all pre-2023 cycles, and comparing only the earliest and latest cycles all produce null results. All specifications include the full set of confounders. Significance levels: $^{*}\,p<0.05$, $^{**}\,p<0.01$, $^{***}\,p<0.001$.
\end{minipage}
\end{table}

\subsection{Check 4. Covariate Stability}
To assess whether the DiD estimate is sensitive to model specification, we estimate the fee waiver $\times$ post interaction across progressively richer covariate sets. Table~\ref{tab:covariate_robustness} reports the results. The baseline specification with no controls yields an interaction coefficient of $\hat{\beta} = -0.195$ (SE $= 0.073$). Adding the first four confounders (i.e. Sex, First Gen, Cumulative GPA, SAT Reading and Writing) attenuates the estimate slightly to $\hat{\beta} = -0.166$ (SE $= 0.075$), and the full model with all confounders produces $\hat{\beta} = -0.170$ (SE $= 0.076$). The total coefficient range across specifications is 0.029, indicating that the estimate is stable and not an artifact of a particular covariate configuration.

However, this stability should be interpreted alongside the COVID sensitivity analysis reported in Table~\ref{tab:covariate_robustness}. While the interaction coefficient is robust to the inclusion or exclusion of covariates within a given sample, it is sensitive to pre-period sample composition: the estimate is driven primarily by the inclusion of pre-2023 cycles rather than by a discrete post-treatment shift. Covariate robustness establishes that no single control variable is responsible for the observed association, but it does not resolve the identification concern raised by the event study and COVID sensitivity analyses. Together, these diagnostics support a descriptive interpretation of the estimates.

\begin{table}[htbp]
\centering
\caption{Covariate Robustness: Fee Waiver $\times$ Post Interaction Across Specifications (Logit)}
\label{tab:covariate_robustness}
\begin{tabular}{lcccccc}
\toprule
\textbf{Specification} & \textbf{Coefficient} & \textbf{Std.\ Error} & \textbf{95\% CI} & \textbf{$p$-value} & \textbf{Pseudo-$R^2$} & \textbf{$N$} \\
\midrule
No controls & $-0.195$ & $(0.073)$ & $[-0.339,\; -0.051]$ & $0.008$ & $0.021$ & $30{,}488$ \\
First 4 confounders & $-0.166$ & $(0.075)$ & $[-0.312,\; -0.019]$ & $0.027$ & $0.066$ & $30{,}488$ \\
All confounders (full) & $-0.170$ & $(0.076)$ & $[-0.318,\; -0.021]$ & $0.025$ & $0.096$ & $30{,}488$ \\
\midrule
\multicolumn{7}{l}{Coefficient range across specifications: $0.029$} \\
\bottomrule
\end{tabular}

\vspace{0.5em}
\begin{minipage}{0.95\textwidth}
\footnotesize
\textit{Notes:} Table reports the coefficient on the fee waiver $\times$ post interaction from logit models with progressively richer covariate sets. The dependent variable is admission. The first four confounders specification includes Sex, First Gen, Cumulative GPA, SAT R/W score. The coefficient is stable across specifications, with a total range of $0.029$. Significance levels: $^{*}\,p<0.05$, $^{**}\,p<0.01$, $^{***}\,p<0.001$.
\end{minipage}
\end{table}

Overall, the aggregate DiD estimate reflects a relative decline in fee-waiver applicant outcomes between the 2022 and 2023 admission cycles, plausibly attributable to the reversion of pandemic-era test-optional policies that had temporarily expanded access for lower-income applicants. The transition directly corresponding to ChatGPT's introduction (2023 $\rightarrow$ 2024) produces no detectable differential effect. We therefore interpret our difference-in-differences estimates as descriptive of shifting associations across admission cycles rather than as evidence that LLM adoption causally affected the relationship between socioeconomic status and admission outcomes. While the absence of a detectable effect does not rule out the possibility that LLM access influenced application quality in ways not captured by admission decisions alone, the current evidence does not support a causal claim.

\section{Stratified Associations between LLM Usage and Admission Probability across SES Groups}\label{appdx:stratified associations}

Table \ref{tab:stratified_effects} reports full results from stratified logistic regression models estimating the association between estimated LLM usage and admission probability separately by SES status. Both models include the same set of controls used in the main specification: cumulative GPA, standardized test scores, sex, first-generation status, school type, and leadership participation.

\begin{table}[!htpb]
    \centering
    \caption{Stratified Associations of LLM Usage on Admission Probability by SES Status}
    \label{tab:stratified_effects}
    \begin{threeparttable}
    \begin{tabular}{l c c c}
        \toprule
        \textbf{Variable} & \textbf{higher SES} & \textbf{lower SES} & \textbf{Difference} \\
         & \textbf{(No Fee Waiver)} & \textbf{(Fee Waiver)} & \textbf{($\Delta$)} \\
        \midrule
        \multicolumn{4}{l}{\textit{Key Variables}} \\
        $\hat{\alpha}$ (LLM Usage) & & & \\
        \hspace{1em} Coefficient & $-0.959$*** & $-1.778$*** & $-0.819$ \\
        \hspace{1em} Standard Error & (0.209) & (0.408) & \\
        \hspace{1em} Odds Ratio [95\% CI] & 0.383 [0.254, 0.577] & 0.169 [0.076, 0.375] & \\
        \hspace{1em} P-value & $<$0.001 & $<$0.001 & \\
        \addlinespace
        \multicolumn{4}{l}{\textit{Control Variables (Selected)}} \\
        Cumulative GPA (scaled) & 0.987*** & 1.511*** & $+0.524$ \\
         & (0.083) & (0.207) & \\
        SAT Reading/Writing & 0.002** & 0.002* & $0.000$ \\
         & (0.001) & (0.001) & \\
        ACT Composite Score & $-0.126$*** & $-0.127$* & $-0.001$ \\
         & (0.029) & (0.058) & \\
        Leadership/Honors & 0.807*** & 0.759*** & $-0.048$ \\
         & (0.078) & (0.118) & \\
        \addlinespace
        \multicolumn{4}{l}{\textit{Model Statistics}} \\
        Observations & 12,193 & 5,461 & \\
        Pseudo $R^2$ & 0.080 & 0.146 & \\
        \bottomrule
    \end{tabular}
    \begin{tablenotes}[flushleft]
        \footnotesize
        \item \textit{Note:} Standard errors in parentheses. Significance levels: *$p < 0.05$, **$p < 0.01$, ***$p < 0.001$.
        \item Models control for sex, first-generation status, school type, GPA, SAT/ACT scores, and leadership/honors.
        \item $\hat{alpha}$ ranges from 0 (no LLM usage detected) to 1 (fully LLM-generated).
        \item The coefficient represents the per unit increase in  $\hat{alpha}$ on log-odds of admission.
        \item Difference ($\Delta$) = lower SES coefficient minus higher SES coefficient. 
    \end{tablenotes}
    \end{threeparttable}
\end{table}

\clearpage
\section{Mediation Analysis}\label{appdx:mediation}

\begin{longtable}{lcccc}
    \caption{Full single-mediator mediation analysis results for each stylometric feature. All estimates are based on 1{,}000 quasi-Bayesian simulations with robust standard errors ($N = 17{,}654$)} \\
    \toprule
    & \textbf{Estimate} &\textbf{95\% CI Lower} &\textbf{95\% CI Upper} & \textbf{$p$-value} \\
    \midrule
    \endfirsthead

    \toprule
    & Estimate & 95\% CI Lower & 95\% CI Upper & $p$-value \\
    \midrule
    \endhead

    \multicolumn{5}{l}{\textbf{Panel A: Mediator = \texttt{n\_tokens} (Token Count)}} \\
    ACME (control)            & $-0.0339$ & $-0.0435$ & $-0.0249$ & $<0.001$ \\
    ACME (treated)            & $-0.0167$ & $-0.0248$ & $-0.0103$ & $<0.001$ \\
    ADE (control)             & $-0.1173$ & $-0.1513$ & $-0.0800$ & $<0.001$ \\
    ADE (treated)             & $-0.1000$ & $-0.1296$ & $-0.0661$ & $<0.001$ \\
    Total Effect              & $-0.1340$ & $-0.1655$ & $-0.0984$ & $<0.001$ \\
    Prop. Mediated (control)  & $0.2525$  & $0.1800$  & $0.3625$  & $<0.001$ \\
    Prop. Mediated (treated)  & $0.1226$  & $0.0703$  & $0.2175$  & $<0.001$ \\
    ACME (average)            & $-0.0253$ & $-0.0336$ & $-0.0180$ & $<0.001$ \\
    ADE (average)             & $-0.1086$ & $-0.1405$ & $-0.0728$ & $<0.001$ \\
    Prop. Mediated (average)  & $0.1875$  & $0.1255$  & $0.2887$  & $<0.001$ \\
    \addlinespace
    
    \multicolumn{5}{l}{\textbf{Panel B: Mediator = \texttt{n\_words} (Word Count)}} \\
    ACME (control)            & $-0.0269$ & $-0.0349$ & $-0.0188$ & $<0.001$ \\
    ACME (treated)            & $-0.0122$ & $-0.0182$ & $-0.0070$ & $<0.001$ \\
    ADE (control)             & $-0.1218$ & $-0.1541$ & $-0.0837$ & $<0.001$ \\
    ADE (treated)             & $-0.1071$ & $-0.1365$ & $-0.0708$ & $<0.001$ \\
    Total Effect              & $-0.1340$ & $-0.1631$ & $-0.0970$ & $<0.001$ \\
    Prop. Mediated (control)  & $0.1987$  & $0.1394$  & $0.2928$  & $<0.001$ \\
    Prop. Mediated (treated)  & $0.0902$  & $0.0485$  & $0.1630$  & $<0.001$ \\
    ACME (average)            & $-0.0195$ & $-0.0262$ & $-0.0132$ & $<0.001$ \\
    ADE (average)             & $-0.1145$ & $-0.1454$ & $-0.0776$ & $<0.001$ \\
    Prop. Mediated (average)  & $0.1444$  & $0.0942$  & $0.2235$  & $<0.001$ \\
    \addlinespace
    
    \multicolumn{5}{l}{\textbf{Panel C: Mediator = \texttt{n\_types} (Unique Word Types)}} \\
    ACME (control)            & $0.0396$  & $0.0299$  & $0.0499$  & $<0.001$ \\
    ACME (treated)            & $0.0209$  & $0.0139$  & $0.0287$  & $<0.001$ \\
    ADE (control)             & $-0.1609$ & $-0.1856$ & $-0.1322$ & $<0.001$ \\
    ADE (treated)             & $-0.1796$ & $-0.2105$ & $-0.1451$ & $<0.001$ \\
    Total Effect              & $-0.1401$ & $-0.1700$ & $-0.1054$ & $<0.001$ \\
    Prop. Mediated (control)  & $-0.2809$ & $-0.4104$ & $-0.1987$ & $<0.001$ \\
    Prop. Mediated (treated)  & $-0.1467$ & $-0.2744$ & $-0.0835$ & $<0.001$ \\
    ACME (average)            & $0.0302$  & $0.0228$  & $0.0386$  & $<0.001$ \\
    ADE (average)             & $-0.1703$ & $-0.1979$ & $-0.1388$ & $<0.001$ \\
    Prop. Mediated (average)  & $-0.2138$ & $-0.3395$ & $-0.1444$ & $<0.001$ \\
    \addlinespace
    
    \multicolumn{5}{l}{\textbf{Panel D: Mediator = \texttt{avg\_word\_len} (Average Word Length)}} \\
    ACME (control)            & $0.2014$  & $0.1562$  & $0.2483$  & $<0.001$ \\
    ACME (treated)            & $0.0607$  & $0.0447$  & $0.0808$  & $<0.001$ \\
    ADE (control)             & $-0.1956$ & $-0.2146$ & $-0.1741$ & $<0.001$ \\
    ADE (treated)             & $-0.3363$ & $-0.3966$ & $-0.2757$ & $<0.001$ \\
    Total Effect              & $-0.1349$ & $-0.1680$ & $-0.0973$ & $<0.001$ \\
    Prop. Mediated (control)  & $-1.4816$ & $-2.1430$ & $-1.0995$ & $<0.001$ \\
    Prop. Mediated (treated)  & $-0.4404$ & $-0.8058$ & $-0.2710$ & $<0.001$ \\
    ACME (average)            & $0.1310$  & $0.1037$  & $0.1587$  & $<0.001$ \\
    ADE (average)             & $-0.2659$ & $-0.3042$ & $-0.2265$ & $<0.001$ \\
    Prop. Mediated (average)  & $-0.9610$ & $-1.4696$ & $-0.7039$ & $<0.001$ \\
    \addlinespace
    
    \multicolumn{5}{l}{\textbf{Panel E: Mediator = \texttt{avg\_sent\_len\_words} (Average Sentence Length)}} \\
    ACME (control)            & $0.0010$  & $-0.0077$ & $0.0103$  & $0.820$ \\
    ACME (treated)            & $0.0006$  & $-0.0047$ & $0.0059$  & $0.820$ \\
    ADE (control)             & $-0.1285$ & $-0.1589$ & $-0.0944$ & $<0.001$ \\
    ADE (treated)             & $-0.1290$ & $-0.1600$ & $-0.0931$ & $<0.001$ \\
    Total Effect              & $-0.1279$ & $-0.1579$ & $-0.0937$ & $<0.001$ \\
    Prop. Mediated (control)  & $-0.0081$ & $-0.0864$ & $0.0618$  & $0.820$ \\
    Prop. Mediated (treated)  & $-0.0043$ & $-0.0537$ & $0.0387$  & $0.820$ \\
    ACME (average)            & $0.0008$  & $-0.0062$ & $0.0081$  & $0.820$ \\
    ADE (average)             & $-0.1287$ & $-0.1592$ & $-0.0936$ & $<0.001$ \\
    Prop. Mediated (average)  & $-0.0062$ & $-0.0683$ & $0.0500$  & $0.820$ \\
    \addlinespace
    
    \multicolumn{5}{l}{\textbf{Panel F: Mediator = \texttt{ttr} (Type-Token Ratio)}} \\
    ACME (control)            & $0.0632$  & $0.0504$  & $0.0768$  & $<0.001$ \\
    ACME (treated)            & $0.0274$  & $0.0198$  & $0.0362$  & $<0.001$ \\
    ADE (control)             & $-0.1615$ & $-0.1873$ & $-0.1337$ & $<0.001$ \\
    ADE (treated)             & $-0.1972$ & $-0.2339$ & $-0.1589$ & $<0.001$ \\
    Total Effect              & $-0.1340$ & $-0.1655$ & $-0.1006$ & $<0.001$ \\
    Prop. Mediated (control)  & $-0.4661$ & $-0.6589$ & $-0.3542$ & $<0.001$ \\
    Prop. Mediated (treated)  & $-0.2005$ & $-0.3464$ & $-0.1245$ & $<0.001$ \\
    ACME (average)            & $0.0453$  & $0.0364$  & $0.0553$  & $<0.001$ \\
    ADE (average)             & $-0.1793$ & $-0.2106$ & $-0.1464$ & $<0.001$ \\
    Prop. Mediated (average)  & $-0.3333$ & $-0.4975$ & $-0.2394$ & $<0.001$ \\
    \addlinespace
    
    \multicolumn{5}{l}{\textbf{Panel G: Mediator = \texttt{maas\_ttr} (Maas TTR)}} \\
    ACME (control)            & $0.0718$  & $0.0594$  & $0.0851$  & $<0.001$ \\
    ACME (treated)            & $0.0309$  & $0.0225$  & $0.0412$  & $<0.001$ \\
    ADE (control)             & $-0.1670$ & $-0.1917$ & $-0.1392$ & $<0.001$ \\
    ADE (treated)             & $-0.2080$ & $-0.2419$ & $-0.1700$ & $<0.001$ \\
    Total Effect              & $-0.1361$ & $-0.1672$ & $-0.1008$ & $<0.001$ \\
    Prop. Mediated (control)  & $-0.5251$ & $-0.7496$ & $-0.4050$ & $<0.001$ \\
    Prop. Mediated (treated)  & $-0.2240$ & $-0.4061$ & $-0.1386$ & $<0.001$ \\
    ACME (average)            & $0.0514$  & $0.0423$  & $0.0608$  & $<0.001$ \\
    ADE (average)             & $-0.1875$ & $-0.2163$ & $-0.1550$ & $<0.001$ \\
    Prop. Mediated (average)  & $-0.3745$ & $-0.5770$ & $-0.2739$ & $<0.001$ \\
    \addlinespace
    
    \multicolumn{5}{l}{\textbf{Panel H: Mediator = \texttt{mtld} (MTLD)}} \\
    ACME (control)            & $0.0522$  & $0.0416$  & $0.0630$  & $<0.001$ \\
    ACME (treated)            & $0.0236$  & $0.0173$  & $0.0314$  & $<0.001$ \\
    ADE (control)             & $-0.1574$ & $-0.1819$ & $-0.1274$ & $<0.001$ \\
    ADE (treated)             & $-0.1860$ & $-0.2198$ & $-0.1485$ & $<0.001$ \\
    Total Effect              & $-0.1338$ & $-0.1634$ & $-0.0984$ & $<0.001$ \\
    Prop. Mediated (control)  & $-0.3875$ & $-0.5499$ & $-0.2943$ & $<0.001$ \\
    Prop. Mediated (treated)  & $-0.1724$ & $-0.3075$ & $-0.1080$ & $<0.001$ \\
    ACME (average)            & $0.0379$  & $0.0305$  & $0.0457$  & $<0.001$ \\
    ADE (average)             & $-0.1717$ & $-0.2014$ & $-0.1381$ & $<0.001$ \\
    Prop. Mediated (average)  & $-0.2800$ & $-0.4273$ & $-0.2032$ & $<0.001$ \\
    \addlinespace
    
    \multicolumn{5}{l}{\textbf{Panel I: Mediator = \texttt{hdd} (HD-D)}} \\
    ACME (control)            & $0.0186$  & $0.0135$  & $0.0240$  & $<0.001$ \\
    ACME (treated)            & $0.0105$  & $0.0071$  & $0.0147$  & $<0.001$ \\
    ADE (control)             & $-0.1425$ & $-0.1710$ & $-0.1102$ & $<0.001$ \\
    ADE (treated)             & $-0.1506$ & $-0.1822$ & $-0.1148$ & $<0.001$ \\
    Total Effect              & $-0.1320$ & $-0.1635$ & $-0.0977$ & $<0.001$ \\
    Prop. Mediated (control)  & $-0.1401$ & $-0.2069$ & $-0.0978$ & $<0.001$ \\
    Prop. Mediated (treated)  & $-0.0783$ & $-0.1395$ & $-0.0460$ & $<0.001$ \\
    ACME (average)            & $0.0145$  & $0.0107$  & $0.0190$  & $<0.001$ \\
    ADE (average)             & $-0.1466$ & $-0.1772$ & $-0.1123$ & $<0.001$ \\
    Prop. Mediated (average)  & $-0.1092$ & $-0.1714$ & $-0.0727$ & $<0.001$ \\
    \addlinespace
    
    \multicolumn{5}{l}{\textbf{Panel J: Mediator = \texttt{yules\_k} (Yule's K)}} \\
    ACME (control)            & $0.0118$  & $0.0079$  & $0.0166$  & $<0.001$ \\
    ACME (treated)            & $0.0069$  & $0.0043$  & $0.0102$  & $<0.001$ \\
    ADE (control)             & $-0.1376$ & $-0.1669$ & $-0.1047$ & $<0.001$ \\
    ADE (treated)             & $-0.1425$ & $-0.1735$ & $-0.1084$ & $<0.001$ \\
    Total Effect              & $-0.1307$ & $-0.1618$ & $-0.0957$ & $<0.001$ \\
    Prop. Mediated (control)  & $-0.0895$ & $-0.1439$ & $-0.0579$ & $<0.001$ \\
    Prop. Mediated (treated)  & $-0.0516$ & $-0.0982$ & $-0.0290$ & $<0.001$ \\
    ACME (average)            & $0.0094$  & $0.0063$  & $0.0133$  & $<0.001$ \\
    ADE (average)             & $-0.1401$ & $-0.1702$ & $-0.1066$ & $<0.001$ \\
    Prop. Mediated (average)  & $-0.0706$ & $-0.1215$ & $-0.0438$ & $<0.001$ \\
    \addlinespace
    
    \multicolumn{5}{l}{\textbf{Panel K: Mediator = \texttt{complexity} (Syntactic Complexity)}} \\
    ACME (control)            & $0.1252$  & $0.0831$  & $0.1720$  & $<0.001$ \\
    ACME (treated)            & $0.0461$  & $0.0315$  & $0.0635$  & $<0.001$ \\
    ADE (control)             & $-0.1777$ & $-0.2006$ & $-0.1494$ & $<0.001$ \\
    ADE (treated)             & $-0.2568$ & $-0.3149$ & $-0.1989$ & $<0.001$ \\
    Total Effect              & $-0.1315$ & $-0.1632$ & $-0.0956$ & $<0.001$ \\
    Prop. Mediated (control)  & $-0.9500$ & $-1.4631$ & $-0.6075$ & $<0.001$ \\
    Prop. Mediated (treated)  & $-0.3467$ & $-0.6454$ & $-0.2022$ & $<0.001$ \\
    ACME (average)            & $0.0857$  & $0.0590$  & $0.1146$  & $<0.001$ \\
    ADE (average)             & $-0.2172$ & $-0.2570$ & $-0.1756$ & $<0.001$ \\
    Prop. Mediated (average)  & $-0.6484$ & $-1.0571$ & $-0.4138$ & $<0.001$ \\
    \addlinespace
    \bottomrule
    \end{longtable}

\end{document}